\documentclass[english,journal]{ieeeaccess}
\usepackage[T1]{fontenc}
\usepackage[latin1]{inputenc}
\usepackage{amsmath}
\usepackage{graphicx}
\usepackage{amssymb}

\makeatletter

\providecommand{\tabularnewline}{\\}

\usepackage{amsmath,amssymb,amsfonts}
\usepackage{algorithmic}

\def\BibTeX{{\rm B\kern-.05em{\sc i\kern-.025em b}\kern-.08em
    T\kern-.1667em\lower.7ex\hbox{E}\kern-.125emX}}

\usepackage{bm}

\usepackage{caption,setspace}
\usepackage{tabto}

\usepackage{babel}
\makeatother
\begin{document}
\doi{10.1109/ACCESS.2023.3348240}

\title{Spline-Shaped Microstrip Edge-Fed Antenna for 77 GHz Automotive Radar Systems}

\author{\uppercase{Marco Salucci}\authorrefmark{1}, \IEEEmembership{Senior Member, IEEE}, \uppercase{Lorenzo Poli}\authorrefmark{1}, \IEEEmembership{Senior Member, IEEE}, \uppercase{Paolo Rocca}\authorrefmark{1,2}, \IEEEmembership{Fellow, IEEE}, \uppercase{Claudio Massagrande}\authorrefmark{3}, \uppercase{Pietro Rosatti}\authorrefmark{1}, \uppercase{Mohammad Abdul Hannan}\authorrefmark{1,4}, \uppercase{Mirko Facchinelli}\authorrefmark{1}, AND \uppercase{Andrea Massa}\authorrefmark{5,1,6,7}, \IEEEmembership{Fellow, IEEE}} 

\address[1]{ELEDIA Research Center (ELEDIA Research Center (ELEDIA@UniTN - University of Trento), DICAM - Department of Civil, Environmental, and Mechanical Engineering, Via Mesiano 77, 38123 Trento - Italy (e-mail: \{marco.salucci, lorenzo.poli, paolo.rocca, pietro.rosatti, mohammadabdul.hannan, mirko.facchinelli, andrea.massa\}@unitn.it)}

\address[2]{ELEDIA Research Center (ELEDIA@XIDIAN - Xidian University), P.O. Box 191, No.2 South Tabai Road, 710071 Xi'an, Shaanxi Province - China (e-mail: paolo.rocca@xidian.edu.cn)}\address[3]{Huawei Technologies Italia, Centro Direzionale Milano 2, Palazzo Verrocchio, 20054 Segrate - Italy  (e-mail: claudio.massagrande1@huawei.com)}\address[4]{ELEDIA Research Center (ELEDIA@UniCT - University of Catania), Department of Electrical, Electronic, and Computer Engineering Via S. Sofia 64, 95123 Catania - Italy (e-mail: mohammadabdul.hannan@unict.it)}\address[5]{ELEDIA Research Center (ELEDIA@UESTC - UESTC), School of Electronic Science and Engineering, University of Electronic Science and Technology of China, Chengdu 611731 - China (e-mail: andrea.massa@uestc.edu.cn)}\address[6]{ELEDIA Research Center (ELEDIA@TSINGHUA - Tsinghua University), 30 Shuangqing Rd, 100084 Haidian, Beijing - China (e-mail: andrea.massa@tsinghua.edu.cn)}

\address[7]{School of Electrical Engineering, Tel Aviv University, Tel Aviv 69978 - Israel (e-mail: andrea.massa@eng.tau.ac.il)}\tfootnote{(c) 2024 IEEE.  Personal use of this material is permitted.  Permission from IEEE must be obtained for all other uses, in any current or future media, including reprinting/republishing this material for advertising or promotional purposes, creating new collective works, for resale or redistribution to servers or lists, or reuse of any copyrighted component of this work in other works. This work benefited from the networking activities carried out within the Project "ICSC National Centre for HPC, Big Data and Quantum Computing (CN HPC)" funded by the European Union - NextGenerationEU within the PNRR Program (CUP: E63C22000970007), the  Project "INSIDE-NEXT - Indoor Smart Illuminator for Device Energization and Next-Generation Communications" funded by the Italian Ministry for Universities and Research within the PRIN 2022 Program (CUP: E53D23000990001), the Project "AURORA - Smart Materials for Ubiquitous Energy Harvesting, Storage, and Delivery in Next Generation Sustainable Environments" funded by the Italian Ministry for Universities and Research within the PRIN-PNRR 2022 Program, the Project "SPEED" (Grant No. 6721001) funded by National Science Foundation of China under the Chang-Jiang Visiting Professorship Program, and the Project DICAM-EXC (Grant L232/2016) funded by the Italian Ministry of Education, Universities and Research (MUR) within the "Departments of Excellence 2023-2027" Program (CUP: E63C22003880001). The work of M. A. Hannan has been supported by the European Union under the Italian National Recovery and Resilience Plan (NRRP) of NextGenerationEU, partnership on "Telecommunications of the Future", PE0000001 - program "RESTART", Structural Project "ISaCAGE".}

\corresp{Corresponding author: Andrea Massa (e-mail: andrea.massa@unitn.it).}

\begin{abstract}
\noindent An innovative millimeter-wave (\emph{mm}-wave) microstrip
edge-fed antenna (\emph{EFA}) for 77 GHz automotive radars is proposed.
The radiator is designed under the main requirements of having (\emph{i})
an horizontally-polarized pattern and (\emph{ii}) a single-layer layout.
Its contour is modeled with a sinusoidal spline-shaped (\emph{SS})
profile characterized by a reduced number of geometrical descriptors,
but still able to guarantee, thanks to a continuously non-uniform
width, a high flexibility in the modeling for fulfilling challenging
user-defined requirements. The \emph{SS-EFA} descriptors are effectively
and efficiently optimized with a customized implementation of the
System-by-Design (\emph{SbD}) paradigm. The synthesized \emph{EFA}
layout, integrated within a linear arrangement of identical replicas
to account for the integration into the real radar system, exhibits
suitable impedance matching, isolation, polarization purity, and stability
of the beam shaping/pointing within the target band $f_{\min}=76$
{[}GHz{]} $\le$ $f$ $\le$ $f_{\max}=78$ {[}GHz{]}. The experimental
assessment, carried out with a Compact Antenna Test Range (\emph{CATR})
system on a printed circuit board (\emph{PCB})-manufactured prototype,
assesses the reliability of the outcomes from the full-wave (\emph{FW})
simulations as well as the suitability of the synthesized \emph{SS-EFA}
for automotive radars.
\end{abstract}

\titlepgskip=-15pt 

\maketitle

\renewcommand{\figurename}{FIGURE}

\renewcommand{\tablename}{TABLE}

\renewcommand{\refname}{REFERENCES}

\section*{Nomenclature}

\NumTabs{12}

\noindent \emph{ACC} \tab Adaptive Cruise Control.

\noindent \emph{AUT} \tab Antenna Under Test.

\noindent \emph{BDD} \tab Beam Direction Deviation.

\noindent \emph{CATR} \tab Compact Antenna Test Range.

\noindent \emph{CFA} \tab Center-Fed Antenna.

\noindent \emph{DoA} \tab Direction of Arrival.

\noindent \emph{DoF} \tab Degree-of-Freedom.

\noindent \emph{DT} \tab\tab Digital Twin.

\noindent \emph{EFA} \tab Edge-Fed Antenna.

\noindent \emph{FBW} \tab Fractional Bandwidth.

\noindent \emph{FF} \tab\tab Far-Field.

\noindent \emph{FMCW} \tab Frequency Modulated Continuous Wave.

\noindent \emph{FW} \tab\tab Full-Wave.

\noindent \emph{LBE} \tab Learning-by-Examples.

\noindent \emph{LHS} \tab Latin Hypercube Sampling.

\noindent \emph{MC} \tab\tab Mutual Coupling.

\noindent \emph{MIMO} \tab Multiple-Input Multiple-Output.

\noindent \emph{OK} \tab\tab Ordinary Kriging.

\noindent \emph{PCB} \tab Printed Circuit Board.

\noindent \emph{PR} \tab\tab Polarization Ratio.

\noindent \emph{PSO} \tab Particle Swarm Optimizer.

\noindent \emph{RS} \tab\tab Resonant.

\noindent \emph{SbD} \tab System-by-Design.

\noindent \emph{SS} \tab\tab Spline-Shaped.

\noindent \emph{SSE} \tab Solution Space Exploration.

\noindent \emph{SW} \tab\tab Standing Wave.

\noindent \emph{TW} \tab\tab Travelling-Wave.

\section{Introduction\label{sec:Intro}}

\noindent Millimiter-wave (\emph{mm}-wave) radars play an important
role in many modern automotive applications ranging from active safety
driver assistance systems to autonomous driving vehicles \cite{Hash 2012}-\cite{Kim 2019}.
Thanks to the capability to measure the distance, the speed, and the
direction of arrival (\emph{DoA}) of multiple targets with low delays,
they are frequently used for blind-spot detection, collision avoidance,
and emergency brake assistance \cite{Overdevest 2020}\cite{Klinefelter 2021}.
Moreover, automotive radars are a key technology for implementing
active comfort systems featuring a high robustness against environmental
conditions including high temperature, darkness, and bad weather.
For instance, let us consider adaptive cruise control (\emph{ACC})
systems that allow the vehicle to autonomously accelerate or brake
or stop in case of traffic jam to relieve the driver of monotonous
tasks.

In order to continuously sense and monitor the surrounding environment
\cite{Hash 2012}, multiple receive and transmit antennas/channels
are used to implement frequency modulated continuous wave (\emph{FMCW})
multiple-input multiple-output (\emph{MIMO}) radars and 77 {[}GHz{]}
solutions are particularly attractive because of the many advantages
over systems operating at lower frequencies (e.g., 24 {[}GHz{]} band
\cite{Hash 2012}\cite{Kuo 2017}\cite{Yu 2019}). As a matter of
fact, the exploitation of the 77 {[}GHz{]} band allows one to design
smaller antennas with lower volume- and weight-related costs as well
as to reach a higher spatial resolution (thanks to the larger absolute
bandwidth of the antenna system) and more precise \emph{DoA} estimations
\cite{Hash 2012}.

As for the antenna implementation, different technological solutions
have been explored in the last few years including Franklin antennas
\cite{Kuo 2017}, series-fed microstrip arrays \cite{Xu 2017}\cite{Guo 2020},
comb-line arrays \cite{Mosalanejad 2018a}-\cite{Zhang 2011}, lens
antennas \cite{Hallbjorner 2012}, dielectric resonator antennas \cite{Abdel-Wahab 2011},
planar grid arrays \cite{Arnieri 2018}\cite{Mosalanejad 2018b},
leaky-wave antennas \cite{Ettorre 2010}, ceramic-filled cavity resonators
\cite{Bauer 2013}, patch arrays \cite{Yoo 2020}\cite{Wang 2013},
and substrate integrated wave-guides (\emph{SIW}s) \cite{Cheng 2009}-\cite{Xu 2014}.
Series-fed architectures are nowadays a mainstream choice because
of the limited cost, the low profile, the light weight, and the simple
manufacturing/integration in automotive systems. However, the feeding
mechanism is more complex than that of traditional corporate feeding
networks since all the (series-connected) radiating points must be
excited in-phase to afford a well-shaped broadside radiation pattern.
To address such an issue, both center-fed (\emph{CFA}s) \cite{Chopra 2019}\cite{Boskovic 2018}
and edge-fed (\emph{EFA}s) \cite{Xu 2017} antennas have been studied.
The former mitigate the beam tilting (or squint) due to the change
of the electrical lengths between consecutive radiating locations
at different frequencies within the working band, but they may be
unsuitable for a compact integration/connection to the driving pins
of a \emph{FMCW} radar micro-chip. Otherwise, \emph{EFA}s can be more
easily closely-packed in linear array arrangements by means of simpler
routing connections \cite{Xu 2017}. However, the design of \emph{EFA}s,
which can be in turn implemented both as resonant (\emph{RS}) and
travelling-wave (\emph{TW}) structures depending on the absence or
presence of a matched load placed at the termination edge on the opposite
of the feeding point \cite{Yi 2019}, turns out to be more challenging
because of the more difficult control of each excitation phase from
a single feeding point located on one edge. Generally speaking, while
\emph{TW-EFA}s usually enable a more flexible beam shaping of the
far-field (\emph{FF}) pattern, they are also characterized by a larger
beam direction deviation (\emph{BDD}) with frequency and a lower radiation
efficiency because of the energy dissipated in the terminating load
\cite{Yi 2019}.

Concerning the polarization, several designs radiating a vertically-polarized
\emph{FF} pattern have been reported in the recent scientific literature
\cite{Kuo 2017}\cite{Xu 2017}\cite{Arnieri 2018}\cite{Yi 2019}.
However, many automotive applications (including the one considered
here) require an horizontal polarization because of the lower backscattering
from road pavements, resulting in a reduced amount of clutter and
thus allowing a more robust target detection \cite{Yu 2019}\cite{Mosalanejad 2018a}\cite{Cheng 2009}\cite{Dewantari 2019}\cite{Ahmed 2020}-\cite{Yi 2023}. 

Finally, another challenging requirement addressed in this work is
the need for a single-layer layout. As a matter of fact, having a
manufacturing process where both the radiating elements and the routing
connections from/towards the controlling radar chipset can be simultaneously
etched on the same face of a single dielectric substrate is highly
desirable in terms of fabrication simplicity/costs and mechanical
robustness to vibrations \cite{Kuo 2017}\cite{Zhang 2011}\cite{Arnieri 2018}\cite{Yoo 2020}.
Therefore, although providing remarkable radiation/bandwidth features,
available multi-layer solutions are not a viable option since they
are not compliant with such a requirement \cite{Mosalanejad 2018b}\cite{Wang 2013}. 

Following this line of reasoning, this paper presents an innovative
single-layer \emph{RS} microstrip \emph{EFA} radiator with horizontal
polarization for automotive radars in the 77 {[}GHz{]} band. Unlike
state-of-art solutions, the contour of the radiating element is modeled
as a spline-shaped (\emph{SS}) profile \cite{Salucci 2018b}-\cite{Salucci 2022}.
Accordingly, arbitrarily-complex shapes can be yielded by acting on
a limited set of geometric degrees-of-freedom (\emph{DoF}s) to fulfil
several challenging requirements on both bandwidth and radiation features.
As a matter of fact, the flexibility of the proposed modeling approach
enables an accurate control of the feeding phases within the operative
band. Moreover, the continuous non-uniformity of the radiator width
allows one to optimize the elevation \emph{FF} features including
a lowering of the sidelobe level (\emph{SLL}), which is beneficial
for reducing both near range clutter from the road surface and the
multi-path interferences. Furthermore, to yield a robust/reliable
design for a fruitful integration within \emph{FMCW} automotive radars,
the non-negligible material losses occurring in the \emph{mm}-wave
regime as well as the mutual coupling (\emph{MC}) effects when integrating
the antenna in the radar system composed by a linear arrangement of
identical elements are taken into account by recurring to a full-wave
(\emph{FW}) modelling of the structure at hand.

Owing to the complexity of the synthesis problem at hand, the System-by-Design
(\emph{SbD}) \cite{Salucci 2022}-\cite{Rosatti 2022} paradigm has
been applied to solve the arising global optimization problem with
a high computational efficiency. More in detail, the design of the
\emph{SS RS-EFA} (shortly in the following \emph{SS-EFA}) has been
carried out by means of a customized version of the \emph{PSO-OK/C}
method \cite{Massa 2021} based on a {}``low dimension'' representation
of the solution space and relying on the {}``collaboration'' between
a Solution Space Exploration (\emph{SSE}) functional block based on
evolutionary operators \cite{Goudos 2021}-\cite{Rocca 2009} and
a Learning-By-Examples (\emph{LBE})-based digital twin (\emph{DT})
of the \emph{FW} simulator aimed at outputting fast predictions of
the electromagnetic (\emph{EM}) performance of each trial guess solution
generated by the \emph{SSE} \cite{Massa 2018b}.%
\begin{figure}
\begin{center}\includegraphics[%
  width=0.65\columnwidth,
  keepaspectratio]{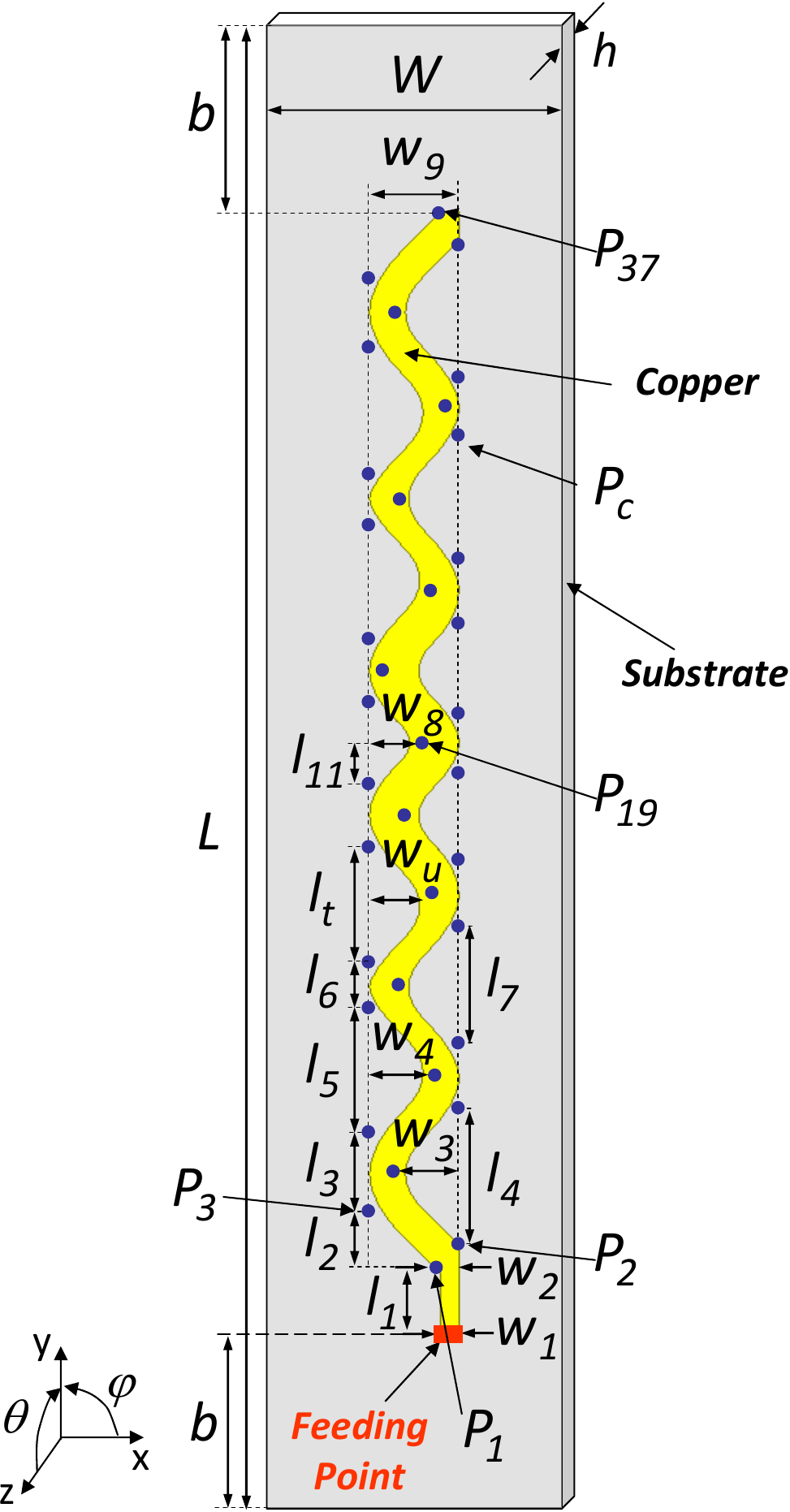}\end{center}

\caption{\emph{SS-EFA} \emph{Design} - Sketch of the antenna geometry with
its geometrical descriptors.}
\end{figure}
It should be noticed that, although other \emph{SS}-based antennas
can be found in the literature \cite{Salucci 2018b}-\cite{Oliveri 2017b},
the present work deals with a completely new design in terms of radiator
architecture, requirements, addressed application, and adopted synthesis
methodology. More in details, \emph{SS} layouts have been previously
solely investigated to model aperture-stacked patches \cite{Salucci 2018b},
coaxial probe-fed microstrip monopoles \cite{Lizzi 2008}, or cavity-backed
patch antennas \cite{Oliveri 2017b} with completely different feeding
mechanisms, \emph{EM} behaviors, and operation bands. Moreover, the
proposed antenna outperforms previous \emph{SS}-based designs \cite{Salucci 2018b}-\cite{Oliveri 2017b}
in terms of beam shaping capabilities and polarization purity, since
it enables a more effective control/tuning of the current density
flowing from the edge feeding point. Furthermore, from the methodological
point of view this is the first time, to the authors' best knowledge,
that the design of a \emph{SS}-based \emph{mm}-wave radiator is addressed
within the \emph{SbD} framework to speed up the optimization process
while considering all non-idealities of the employed materials and
the presence of \emph{MC}. The paper is organized as follows. Section
II describes the \emph{SS-EFA} geometry and it provides some theoretical
insights on its \emph{EM} working principle. The design requirements
and the \emph{SbD} procedure for the optimization of the antenna layout
are detailed in Sect. III. Section IV presents selected results from
the \emph{FW}-based numerical assessment of the performance of the
optimized radiator along with the experimental validation of a \emph{SS-EFA}
prototype realized on a printed circuit board (\emph{PCB}) and measured
on a custom \emph{mm}-wave Compact Antenna Test Range (\emph{CATR})
system. Finally, some conclusions are drawn (Sect. V).

\section{\emph{SS-EFA} Layout and \emph{EM} Working Principle \label{sec:SS-EFA-Layout-and-Working}}

\noindent The layout of the \emph{SS-EFA} is shown in Fig. 1. The
antenna lies on the $\left(x,\, y\right)$ plane and it is printed
on a single-layer ground-backed Rogers $RO3003^{TM}$ high frequency
ceramic-filled composite substrate with relative permittivity and
loss tangent equal to $\varepsilon_{r}=3.0$ and $\tan\delta=1.0\times10^{-3}$,
respectively, of thickness $h=0.127$ {[}mm{]} (i.e., $h=5$ {[}mil{]}).
To take into account all non-idealities/losses occurring in the \emph{mm}-wave
band, both metallizations on the top (i.e., the radiator) and the
bottom (i.e., the ground plane) layers are modeled as copper with
conductivity $\sigma=2.5\times10^{7}$ {[}S/m{]}, thickness $\tau=25\times10^{-6}$
{[}m{]}, and roughness $\rho=1.3\times10^{-6}$ {[}m{]}. %
\begin{figure}
\begin{center}\includegraphics[%
  width=0.90\columnwidth,
  keepaspectratio]{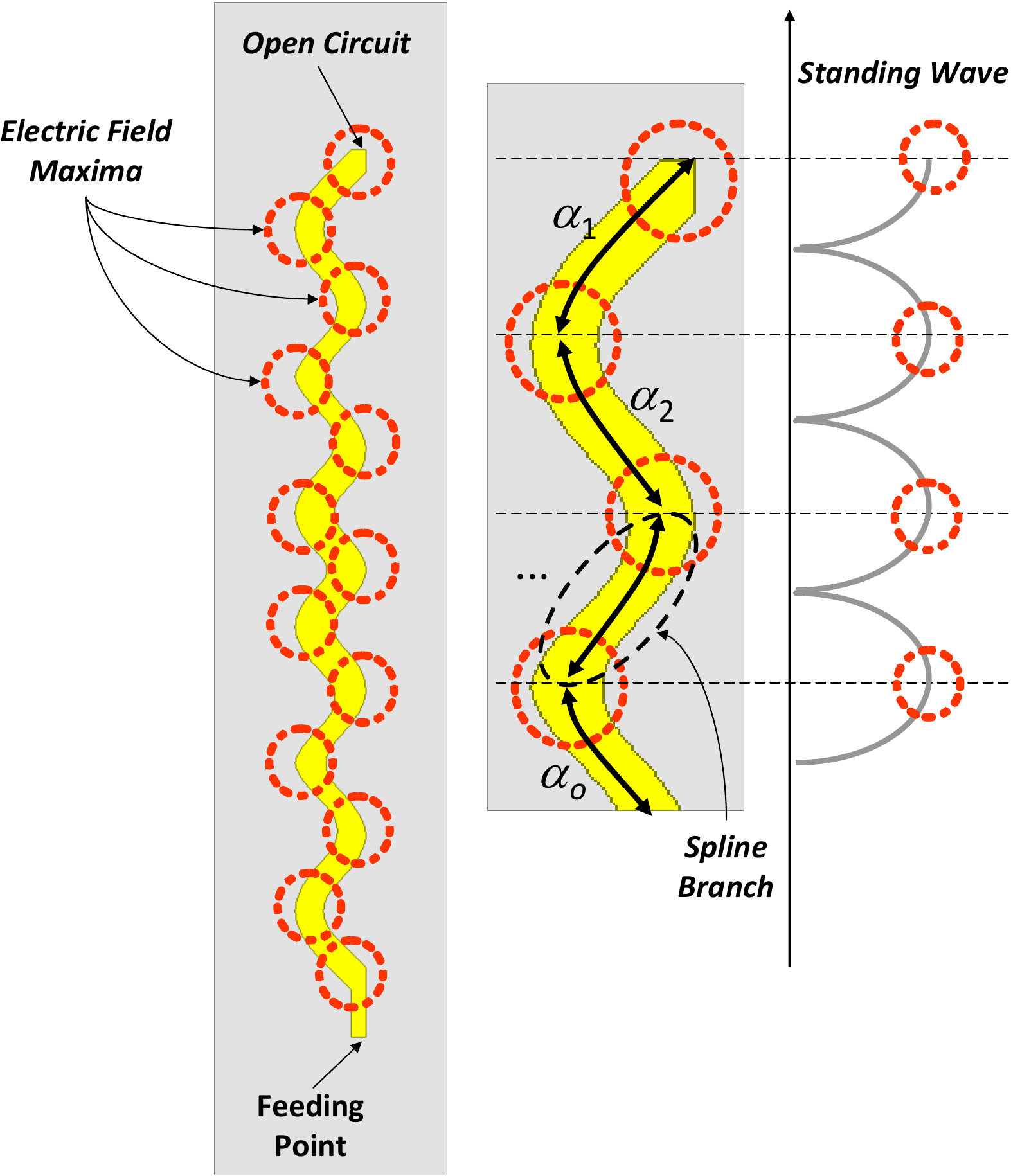}\end{center}

\caption{\emph{SS-EFA} \emph{Working Principle} - Pictorial representation
of the \emph{SW} excited within the \emph{SS-EFA} structure.}
\end{figure}

The antenna is fed from the bottom edge with a tapered microstrip
feeding line of length $l_{1}$, having controllable starting, $w_{1}$,
and ending, $w_{2}$, widths to yield a proper $50$ {[}$\Omega${]}
impedance matching over the complete working band %
\footnote{\noindent In order to maximize the power flow into the \emph{SS} series-fed
radiating structure, a tapered input section is adopted and its geometric
descriptors $w_{1}$ and $w_{2}$ are included in the design variables,
as detailed in the following.%
}. As for the shape of the radiator, which is connected to the feeding
line, an innovative non-uniform \emph{SS} profile is adopted \cite{Salucci 2018b}-\cite{Salucci 2022}
since (\emph{i}) there is the possibility to model complex geometries
for fitting multiple and sometimes contrasting requirements on both
bandwidth and far-field (\emph{FF}) features with a reduced set of
properly-tuned \emph{DoF}s; (\emph{ii}) the absence of sharp edges
is a recipe to mitigate the fringing effects that enhance the \emph{MC}
between the adjacent elements of the final \emph{}radar layout \cite{Salucci 2018b}.
More in detail, Bézier spline basis functions \cite{Salucci 2018b},
with $C=37$ control points, $\underline{P}=\left\{ P_{c}=\left(x_{c},\, y_{c}\right);\, c=1,...,C\right\} $
(Fig. 1), are adopted to shape the contour of the \emph{SS-EFA}. 

The spline radiator is terminated on the opposite side of the feeding
point with an open circuit (i.e., no matched load) so that a resonant
behavior is yielded by exciting a standing wave (\emph{SW}) within
the microstrip structure (Fig. 2). As a matter of fact, the maxima
of the electric field occur in correspondence of the bending corners
of the \emph{SS-EFA} that, in turn, correspond to the positions of
the \emph{SW} maxima (Fig. 2). To afford the desired resonating behavior,
the surface current in each spline branch must be tuned in-phase so
that the \emph{FF} radiated contributions constructively add in the
antenna broadside direction {[}i.e., $\left(\theta_{0},\,\varphi_{0}\right)=\left(0,\,0\right)$
{[}deg{]} - Fig. 1{]}. Towards this end, the electrical length of
each $o$-th ($o=1,...,O$; $O\triangleq\frac{C-1}{3}$ $\to$ $O=12$)
intermediate spline segment, $\alpha_{o}$, must be properly designed
so that the current, injected from the feeding point and flowing up
to the open circuit, is equally-phased at each radiating location
(Fig. 2) despite the single-point edge-feeding mechanism.%
\begin{figure}
\begin{center}\includegraphics[%
  width=0.60\columnwidth,
  keepaspectratio]{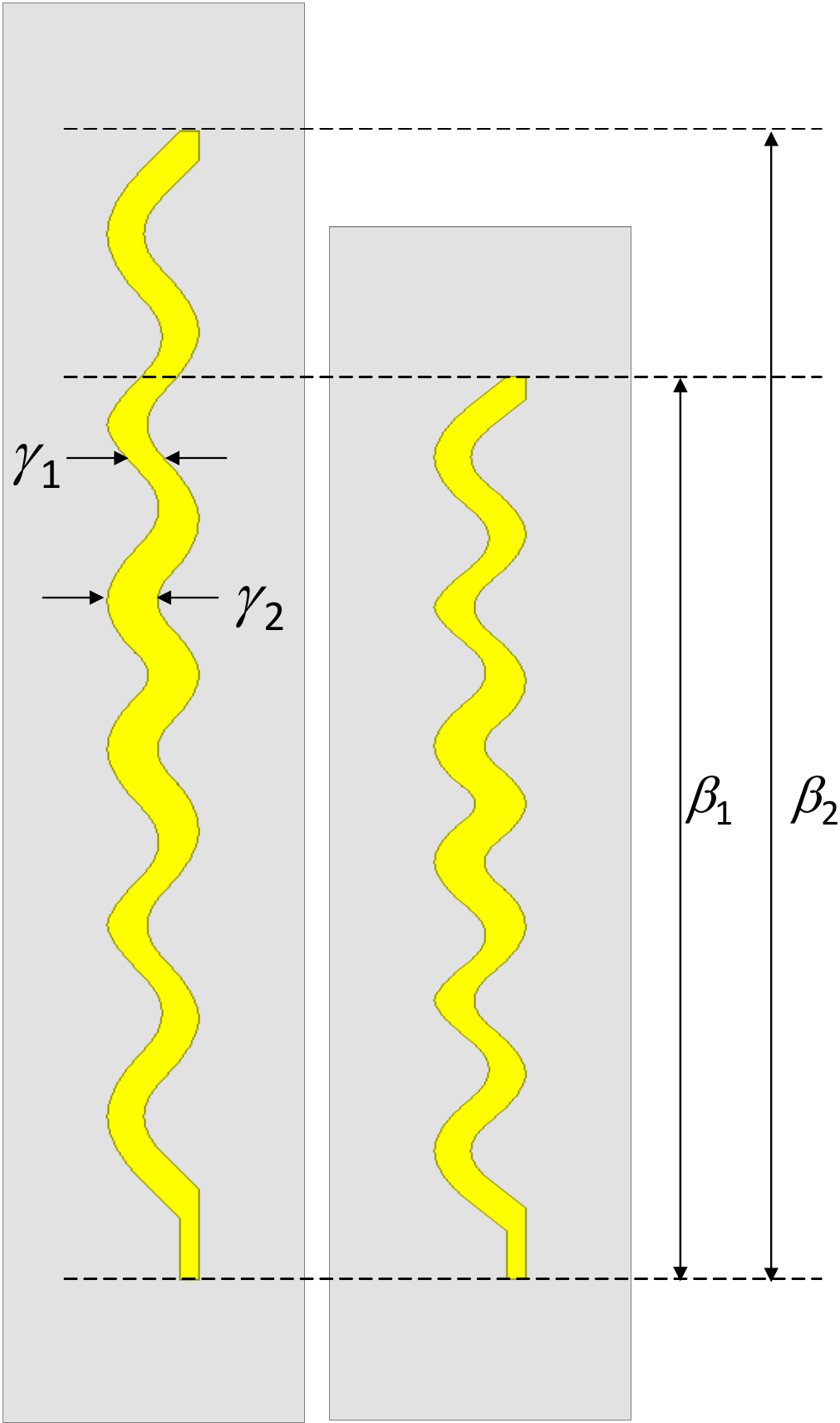}\end{center}

\caption{\emph{SS-EFA} \emph{Working Principle} - Sketch of two non-uniform
width ($\gamma_{1}\neq\gamma_{2}$) and different length ($\beta_{1}\neq\beta_{2}$)
\emph{SS-EFA}s.}
\end{figure}

It is also worth pointing out that thanks to the continuously non-uniform
width of the \emph{SS} metallization (e.g., $\gamma_{1}\neq\gamma_{2}$
- Fig. 3), it is possible to excite a tapered current distribution
within the \emph{EFA} to perform beam shaping and obtain a lower \emph{SLL}
with respect to a uniform-width profile. Such a \emph{FF} feature
is highly desirable in automotive applications since it reduces the
interferences from both the asphalt and the sky, thus leading to a
more robust and reliable target detection/location and \emph{DoA}
estimation. Clearly, the non-uniform width of the spline profile cannot
be arbitrarily set, and its local tuning must be optimized to define
a suitable trade-off between the \emph{SLL} and other important performance
indexes (e.g., impedance matching).

As for the \emph{FF} half-power beamwidth (\emph{HPBW}), which affects
the angular resolution of the automotive radar along the elevation
plane, it is controlled by the overall length of the \emph{SS} radiator.
Indeed, longer structures exhibit narrower \emph{HPBW}s (e.g., $\beta_{2}>\beta_{1}$
$\rightarrow$ $\left.HPBW\right|_{\beta_{2}}<\left.HPBW\right|_{\beta_{1}}$-
Fig. 3) and vice-versa, because of the different aperture size of
the equivalent linear array.

In order to guarantee geometric/electric symmetry by also keeping
low the number of problem descriptors, only one half of the spline
curve (i.e., \{$P_{c}$; $c=1,...,19$\} - Fig. 1) is optimized, while
the coordinates of the second half of the control points (i.e., \{$P_{c}$;
$c=20,...,C$\} - Fig. 1) are automatically derived with a mirroring
operation with respect to the horizontal axis (i.e., $x$-axis - Fig.
1). Enforcing such a symmetry allows one to also yield a lower \emph{BDD}
and a higher similarity of the left/right side-lobes on the vertical
plane (i.e., $\varphi=90$ {[}deg{]} - Fig. 1).%
\begin{figure}
\begin{center}\includegraphics[%
  width=0.60\columnwidth,
  keepaspectratio]{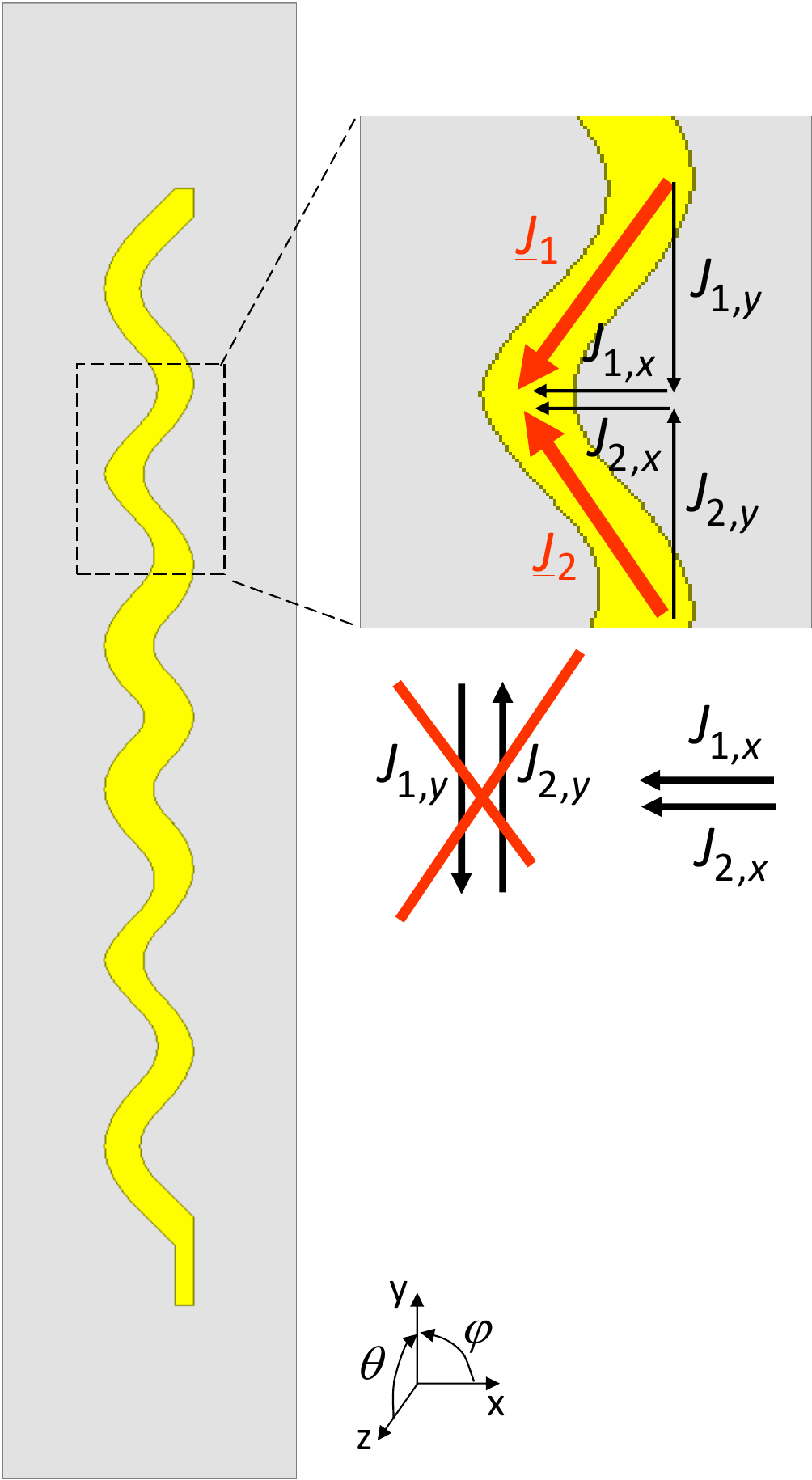}\end{center}

\caption{\emph{SS-EFA} \emph{Working Principle} - Pictorial representation
of the behavior of the surface current density.}
\end{figure}

Finally, it is important to point out that despite the unconventional
and non-uniform shaping of the radiating element, the \emph{SS-EFA}
radiates, as required, a linearly polarized field along the horizontal
direction with a high polarization purity. As a matter of fact, the
surface current density distributions excited within each pair of
adjacent spline branches (e.g., $\underline{J}_{1}=J_{1,x}\underline{\widehat{x}}+J_{1,y}\underline{\widehat{y}}$
and $\underline{J}_{2}=J_{2,x}\underline{\widehat{x}}+J_{2,y}\underline{\widehat{y}}$
- Fig. 4) exhibit in-phase $x$-components (e.g., $J_{1,x}$ and $J_{2,x}$
- Fig. 4) and out-of-phase $y$-components (e.g., $J_{1,y}$ and $J_{2,y}$
- Fig. 4) so that, while the $x$-components constructively sum, the
$y$ ones cancel out and the \emph{EM} source turns out to be $x$-polarized.

\section{Design Process \label{sec:Antenna-Design}}

\noindent The \emph{SS-EFA} has been synthesized to provide a suitable
matching so that $\left|S_{11}\left(f\right)\right|\leq S_{11}^{th}$
($S_{11}^{th}=-10$ {[}dB{]}) within the frequency range $\Delta f$
$=$ $\left[f_{\min},\, f_{\max}\right]$ \cite{Yu 2018}, $f_{\min}$
and $f_{\max}$ being the minimum ($f_{\min}=76$ {[}GHz{]}) and the
maximum ($f_{\max}=78$ {[}GHz{]}) working frequency, respectively.
Concerning the radiation features, the \emph{SLL,} the \emph{HPBW},
and the \emph{BDD} on the vertical plane have been required to comply
with the following requirements: $SLL\left(f\right)\leq SLL^{th}$
($SLL^{th}=-15$ {[}dB{]}), $HPBW\left(f\right)\leq HPBW^{th}$ ($HPBW^{th}=18$
{[}deg{]}), and $BDD\left(f\right)\leq BDD^{th}$ ($BDD^{th}=2$ {[}deg{]})%
\footnote{\noindent The \emph{BDD} is defined as $BDD\left(f\right)=\left|\theta_{\max}\left(f\right)\right|$,
where $\theta_{\max}\left(f\right)=\arg\left\{ \max_{\theta}\left[\left.G\left(f,\,\theta,\,\varphi\right)\right|_{\varphi=90\,[\mathrm{deg}]}\right]\right\} $,
$\left.G\left(f,\,\theta,\,\varphi\right)\right|_{\varphi=90\,[\mathrm{deg}]}$
being the gain pattern function in the vertical (elevation) plane
of the antenna (Fig. 1).%
}. Furthermore, the polarization ratio (\emph{PR}) is required to be
$PR\left(f\right)\geq PR^{th}$ ($PR^{th}=20$ {[}dB{]})%
\footnote{\noindent The \emph{PR} is defined as $PR\left(f\right)=\left|\frac{E_{x}\left(f,\,\theta=0\right)}{E_{y}\left(f,\,\theta=0\right)}\right|$,
$E_{x/y}$ being the $x/y$-components of the \emph{FF} electric field,
respectively.%
}. For the sake of clarity, all design objectives/targets are reported
in Tab. I.%
\begin{table}

\caption{Radar antenna requirements.}

\begin{center}\resizebox{\columnwidth}{!}{\begin{tabular}{|c|c|}
\hline 
\emph{Feature}&
\emph{Requirement}\tabularnewline
\hline
\hline 
Operating Band&
$f\in\left[f_{\min},\, f_{\max}\right]=\left[76,\,78\right]$ {[}GHz{]}\tabularnewline
\hline 
Reflection Coefficient&
$S_{11}\leq S_{11}^{th}=-10$ {[}dB{]}\tabularnewline
\hline 
Sidelobe Level&
$SLL\leq SLL^{th}=-15$ {[}dB{]}\tabularnewline
\hline 
Half-Power Beamwidth&
$HPBW\leq HPBW^{th}=18$ {[}deg{]}\tabularnewline
\hline 
Beam Direction Deviation&
$BDD\leq BDD^{th}=2$ {[}deg{]}\tabularnewline
\hline 
Polarization Ratio&
$PR\geq PR^{th}=20$ {[}dB{]}\tabularnewline
\hline
\end{tabular}}\end{center}
\end{table}

To yield a robust design and to enable a reliable prediction of the
\emph{EM} behavior of the elementary radiator when integrated in an
automotive radar system %
\footnote{\noindent \emph{FMCW} multiple-input multiple-output (\emph{MIMO})
automotive radars are generally implemented as properly-spaced arrangements
of both transmitting and receiving elementary radiators connected
to a single driving chip \cite{Hash 2012}\cite{Kan 2022}\cite{Ahmad 2021}.%
}, it has been synthesized not alone, but within a linear arrangement
of $N=5$ identical half-wavelength ($W=\frac{\lambda_{0}}{2}=1.95\times10^{-3}$
{[}m{]}, $\lambda_{0}$ being the free-space wavelength at the central
frequency $f_{0}=77$ {[}GHz{]}) spaced \emph{SS-EFA}s ($W_{s}=15$
{[}mm{]} - Fig. 5), which has been modeled with a \emph{FW} finite
model. More specifically, the synthesis has been aimed at optimizing
all performance indexes for the central embedded element, while the
surrounding $\left(N-1\right)$ replicas have been terminated on $50$
{[}$\Omega${]} matched loads (Fig. 5).%
\begin{figure}
\begin{center}\includegraphics[%
  width=0.85\columnwidth]{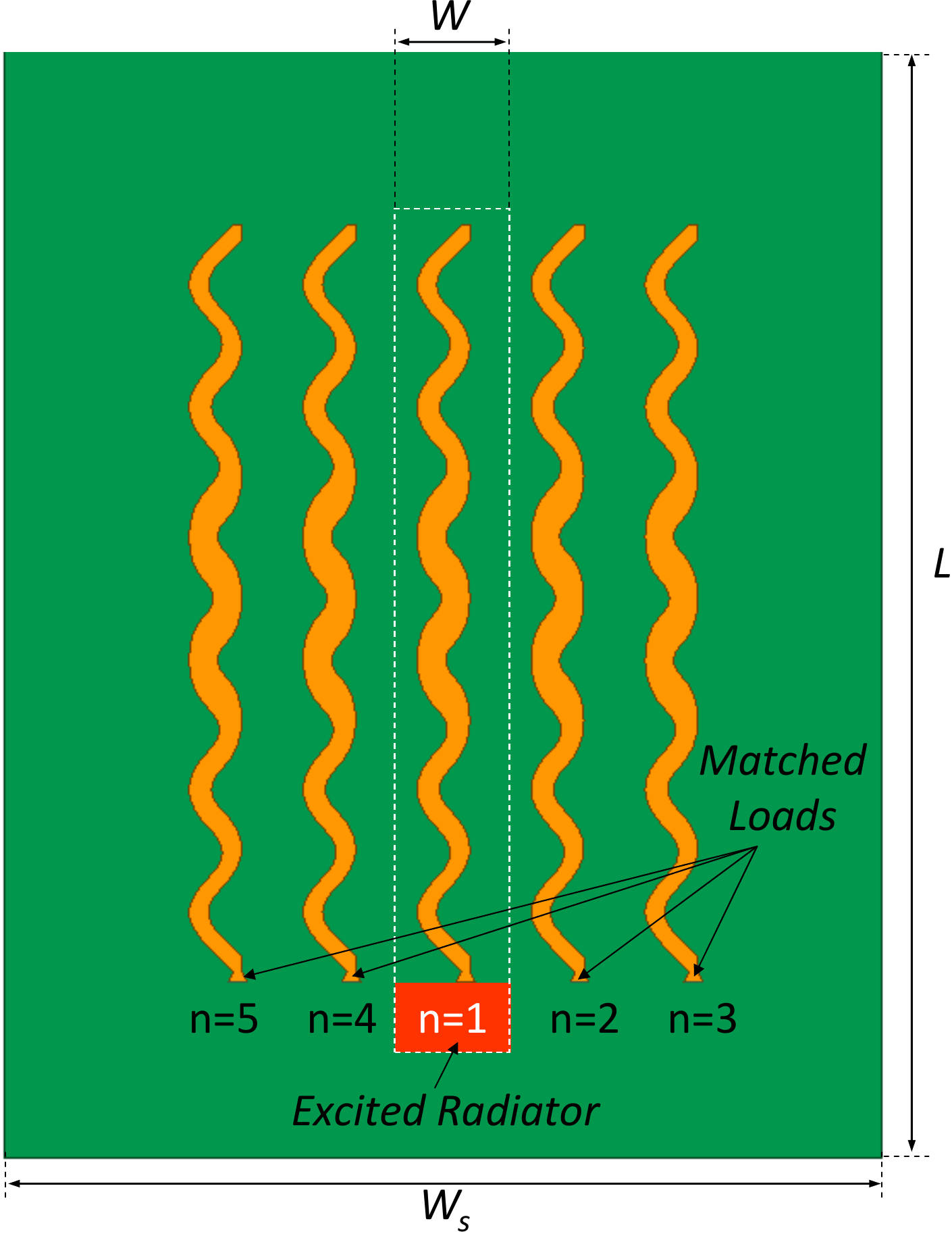}\end{center}

\caption{\emph{Numerical Assessment} ($N=5$) - Screenshot of the \emph{SbD}-synthesized
optimal \emph{SS-EFA} radiator layout embedded within a $N=5$ linear
array of identical $\frac{\lambda_{0}}{2}$-spaced replicas.}
\end{figure}

Owing to the computational complexity of the synthesis problem at
hand, the design has been efficiently carried out within the \emph{SbD}
framework \cite{Salucci 2022}-\cite{Rosatti 2022}. As a matter of
fact, the \emph{SbD} has recently emerged as an innovative paradigm
enabling an effective and computationally-efficient use of global
optimizers for the solution of complex \emph{EM} synthesis problems.
Accordingly, the computational burden (resulting from the need for
iterated accurate \emph{FW}-assessments of the finite structure in
Fig. 5) is addressed by suitably selecting and integrating \emph{functional
blocks} comprising problem-dependent, efficient, and reliable prediction
and optimization strategies \cite{Massa 2021}. More specifically,
the customization of the \emph{SbD} to the synthesis problem at hand
starts from a {}``smart'' representation of the \emph{SS-EFA} solution
space in the so-called \emph{{}``Problem Formulation''} functional
block \cite{Massa 2021}. Towards this end, the following $K=20$
descriptors $\underline{\chi}=\left\{ \chi_{k};\, k=1,...,K\right\} $
where $\chi_{k}=l_{k}$ ($k=1,...,T$; $T=11$) and $\chi_{k}=w_{k-T}$
($k=T+1,...,T+U$; $U=9$) (Fig. 1) have been considered instead of
using the coordinates of the control points of the spline contour
(i.e., $K=2\times C$ $\to$ $K=74$). The synthesis problem has been
then reformulated as a minimization one by defining the following
customized cost function\begin{equation}
\begin{array}{r}
\Phi\left(\underline{\chi}\right)=\alpha_{S_{11}}\Phi_{S_{11}}\left\{ \underline{\chi}\right\} +\alpha_{SLL}\Phi_{SLL}\left\{ \underline{\chi}\right\} +\alpha_{HPBW}\times\\
\times\Phi_{HPBW}\left\{ \underline{\chi}\right\} +\alpha_{BDD}\Phi_{BDD}\left\{ \underline{\chi}\right\} +\alpha_{PR}\Phi_{PR}\left\{ \underline{\chi}\right\} ,\end{array}\label{eq:cost-total}\end{equation}
where\begin{equation}
\Phi_{S_{11}}\left\{ \underline{\chi}\right\} =\frac{1}{Q}\sum_{q=1}^{Q}\mathcal{H}\left\{ \frac{S_{11}\left(f_{q},\,\underline{\chi}\right)-S_{11}^{th}}{\left|S_{11}^{th}\right|}\right\} ,\label{eq:cost_S11}\end{equation}
\begin{equation}
\Phi_{SLL}\left\{ \underline{\chi}\right\} =\frac{1}{Q}\sum_{q=1}^{Q}\mathcal{H}\left\{ \frac{SLL\left(f_{q},\,\underline{\chi}\right)-SLL^{th}}{\left|SLL^{th}\right|}\right\} ,\label{eq:cost_SLL}\end{equation}
\begin{equation}
\Phi_{HPBW}\left\{ \underline{\chi}\right\} =\frac{1}{Q}\sum_{q=1}^{Q}\mathcal{H}\left\{ \frac{HPBW\left(f_{q},\,\underline{\chi}\right)-HPBW^{th}}{HPBW^{th}}\right\} ,\label{eq:cost_HPBW}\end{equation}
\begin{equation}
\Phi_{BDD}\left\{ \underline{\chi}\right\} =\frac{1}{Q}\sum_{q=1}^{Q}\mathcal{H}\left\{ \frac{BDD\left(f_{q},\,\underline{\chi}\right)-BDD^{th}}{BDD^{th}}\right\} ,\label{eq:cost_BDD}\end{equation}
and\begin{equation}
\Phi_{PR}\left\{ \underline{\chi}\right\} =\frac{1}{Q}\sum_{q=1}^{Q}\mathcal{H}\left\{ \frac{PR^{th}-PR\left(f_{q},\,\underline{\chi}\right)}{PR^{th}}\right\} \label{eq:cost_PR}\end{equation}
are the impedance matching, the \emph{SLL}, the \emph{HPBW}, the \emph{BDD},
and the \emph{PR} cost terms quantifying the mismatch with respect
to the corresponding user-defined thresholds. Moreover, $\underline{\alpha}=\left\{ \alpha_{S_{11}};\,\alpha_{SLL};\,\alpha_{HPBW};\,\alpha_{BDD};\,\alpha_{PR}\right\} \geq0$
are real valued weights, while $\mathcal{H}\left\{ \xi\right\} =\xi$
if $\xi\geq0$ and $\mathcal{H}\left\{ \xi\right\} =0$, otherwise,
is the ramp function. Finally, the frequency band $\Delta f$ has
been uniformly sampled into $Q=41$ points, \{$f_{q}=f_{\min}+\left(q-1\right)\frac{\left(f_{\max}-f_{\min}\right)}{\left(Q-1\right)}$;
$q=1,...,Q$\}, for constraining the fulfilment of the requirements
in the whole bandwidth. %
\begin{figure}
\begin{center}\includegraphics[%
  width=0.90\columnwidth]{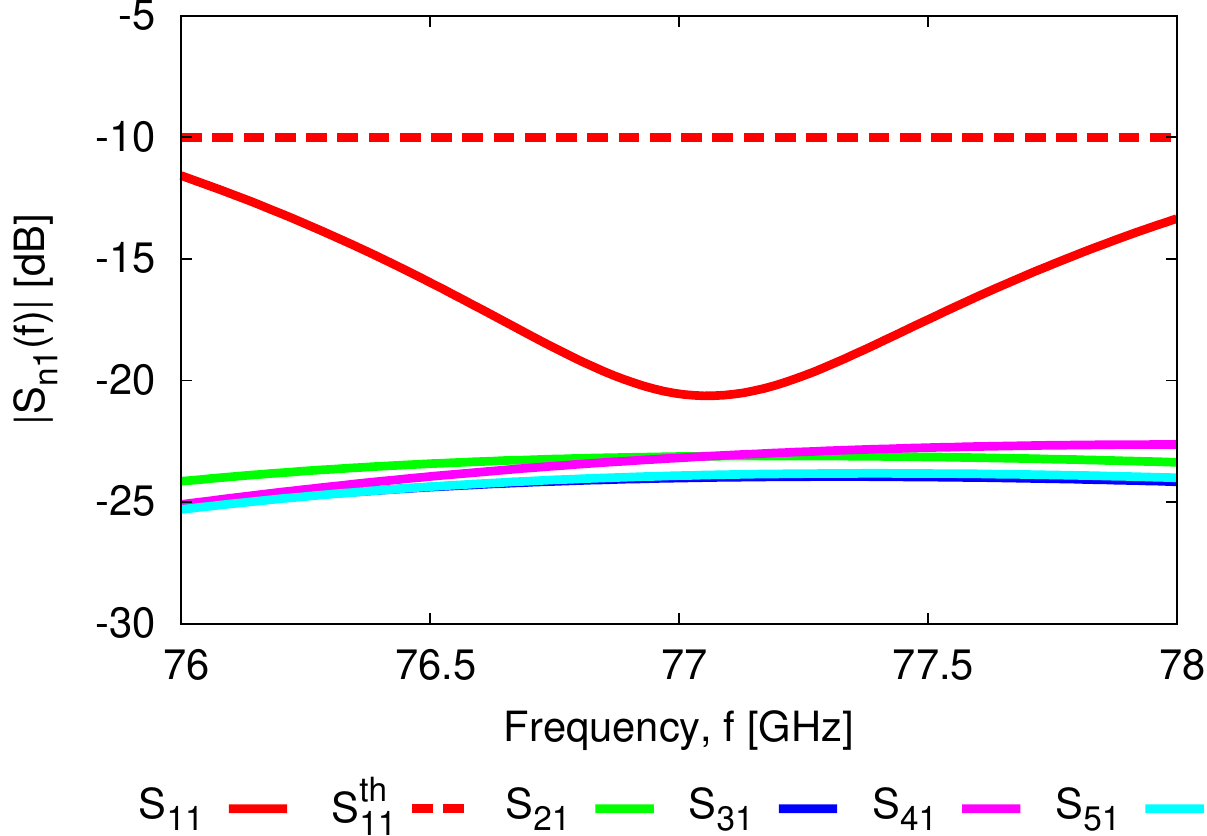}\end{center}

\caption{\emph{Numerical Assessment} ($N=5$) - Magnitude of the reflection
coefficient, $\left|S_{11}\left(f\right)\right|$ ($n=1$), and of
the scattering coefficients, $\left|S_{n1}\left(f\right)\right|$
($n=2,...,N$), of the central embedded element ($n=1$) versus the
frequency $f$ ($f\in\Delta f$) for the linear arrangement in Fig.
5. }
\end{figure}

The multi-modal nature of the cost function (\ref{eq:cost-total})
as well as the impossibility to derive closed-forms expressions of
its derivatives prohibit the exploitation of fast local-search gradient-descent
algorithms. On the other hand, a global optimization using standard
evolutionary algorithms, although guaranteeing an effective exploration
of the solution space without being trapped into local minima, would
result in a very high computational load \cite{Massa 2021}. Owing
to such considerations, the minimization of (\ref{eq:cost-total})
has been carried out with a properly customized version of the \emph{PSO-OK/C}
\emph{SbD} method \cite{Massa 2021} leveraging on the {}``collaboration''
between the \emph{SSE} functional block relying on the Particle Swarm
Optimization (\emph{PSO}) operators \cite{Rocca 2009} and a fast
\emph{DT} based on the Ordinary Kriging (\emph{OK}) \emph{LBE} technique
\cite{Massa 2018b}. The meta-level control parameters of the \emph{PSO}
have been set according to the literature guidelines \cite{Massa 2021}:
$V=10$ ($V$ being the swarm size), $I=200$ ($I$ being the maximum
number of iterations), $\omega=0.4$ ($\omega$ being the inertial
weight), and $\mathcal{C}_{1}=\mathcal{C}_{2}=2.0$ ($\mathcal{C}_{1}$
and $\mathcal{C}_{2}$ being the social and the cognitive acceleration
coefficient, respectively). %
\begin{figure}
\begin{center}\begin{tabular}{c}
\includegraphics[%
  width=0.90\columnwidth]{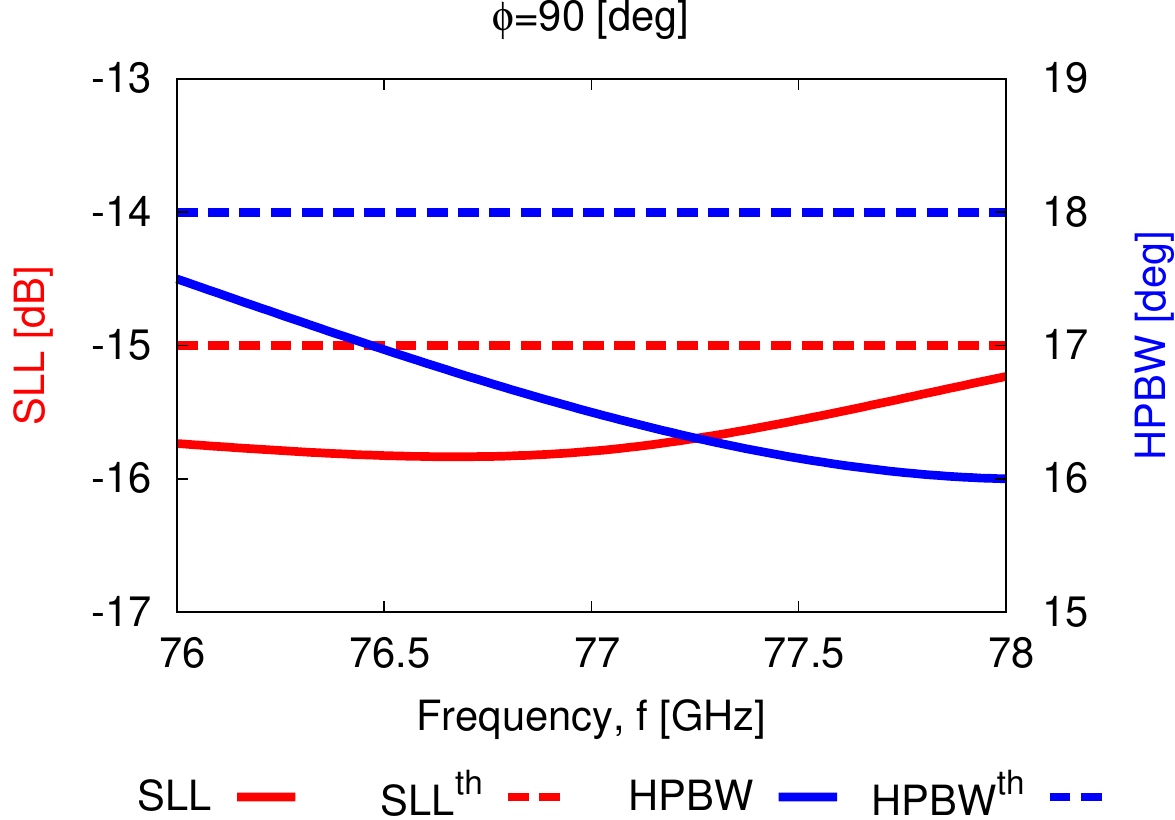}\tabularnewline
(\emph{a})\tabularnewline
\tabularnewline
\includegraphics[%
  width=0.90\columnwidth]{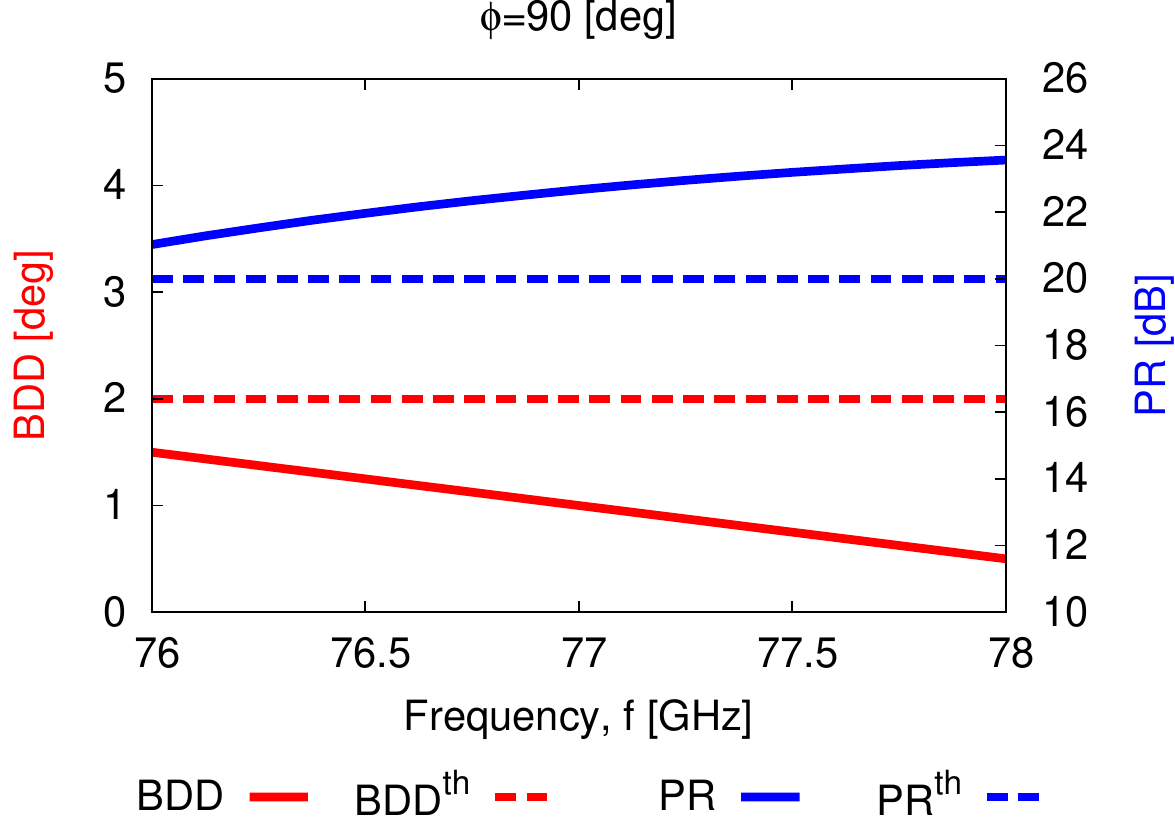}\tabularnewline
(\emph{b})\tabularnewline
\end{tabular}\end{center}

\caption{\emph{Numerical Assessment} ($N=5$, $n=1$, $\varphi=90$ {[}deg{]})
- Behavior of (\emph{a}) \emph{SLL} and \emph{HPBW} and (\emph{b})
\emph{BDD} and \emph{PR} versus the frequency $f$ ($f\in\Delta f$).}
\end{figure}

As for the \emph{OK} prediction model, the fast surrogate of the \emph{FW}
simulator has been trained with $S_{0}=100$ training samples generated
\emph{off-line} according to the Latin Hypercube Sampling (\emph{LHS})
strategy \cite{Massa 2021}, while $S_{upd}=200$ {}``reinforcement
training'' \emph{FW} simulations have been performed \emph{on-line}
by the \emph{PSO-OK/C} to adaptively enhance the \emph{DT} accuracy
during the iterative minimization of (\ref{eq:cost-total}).%
\begin{table}

\caption{Values of the descriptors of the \emph{SbD}-synthesized \emph{SS-EFA}
layout.}

\begin{center}\begin{tabular}{|c|c||c|c|}
\hline 
\multicolumn{4}{|c|}{Geometrical Descriptors {[}m{]}}\tabularnewline
\hline
\hline 
$w_{1}$&
$3.26\times10^{-4}$&
$l_{2}$&
$7.57\times10^{-4}$\tabularnewline
\hline 
$w_{2}$&
$1.46\times10^{-4}$&
$l_{3}$&
$5.27\times10^{-4}$\tabularnewline
\hline 
$w_{3}$&
$7.57\times10^{-4}$&
$l_{4}$&
$1.51\times10^{-3}$\tabularnewline
\hline 
$w_{4}$&
$7.57\times10^{-4}$&
$l_{5}$&
$1.51\times10^{-3}$\tabularnewline
\hline 
$w_{5}$&
$6.90\times10^{-4}$&
$l_{6}$&
$6.47\times10^{-4}$\tabularnewline
\hline 
$w_{6}$&
$6.90\times10^{-4}$&
$l_{7}$&
$1.38\times10^{-3}$\tabularnewline
\hline 
$w_{7}$&
$5.52\times10^{-4}$&
$l_{8}$&
$1.38\times10^{-3}$\tabularnewline
\hline 
$w_{8}$&
$5.52\times10^{-4}$&
$l_{9}$&
$1.00\times10^{-3}$\tabularnewline
\hline 
$w_{9}$&
$9.03\times10^{-4}$&
$l_{10}$&
$1.10\times10^{-3}$\tabularnewline
\hline 
$l_{1}$&
$1.80\times10^{-4}$&
$l_{11}$&
$5.22\times10^{-4}$\tabularnewline
\hline
\end{tabular}\end{center}
\end{table}

The solution, $\underline{\chi}^{\left(opt\right)}$, outputted by
the \emph{SbD} at the convergence {[}i.e., $\Phi\left(\underline{\chi}^{\left(opt\right)}\right)=0${]}
that fulfils all the user-defined requirements (letting $\underline{\alpha}=1.0$),
as it can be inferred in Sect. IV-A, is reported in Tab. II, while
the \emph{CAD} model of the corresponding \emph{SS-EFA} layout (surrounded
by four identical replicas) is shown in Fig. 5, the total length of
the radiator being equal to $L=l_{1}+2\times\left(b+l_{2}+l_{3}+l_{5}+l_{6}+l_{8}+l_{9}+l_{11}\right)$
$\to$ $L=18.9\times10^{-3}$ {[}m{]} since $b=3\times10^{-3}$ {[}m{]}
is a fixed offset from the bottom and the top edges of the substrate
(Fig. 1).%
\begin{figure}
\begin{center}\includegraphics[%
  width=0.60\columnwidth]{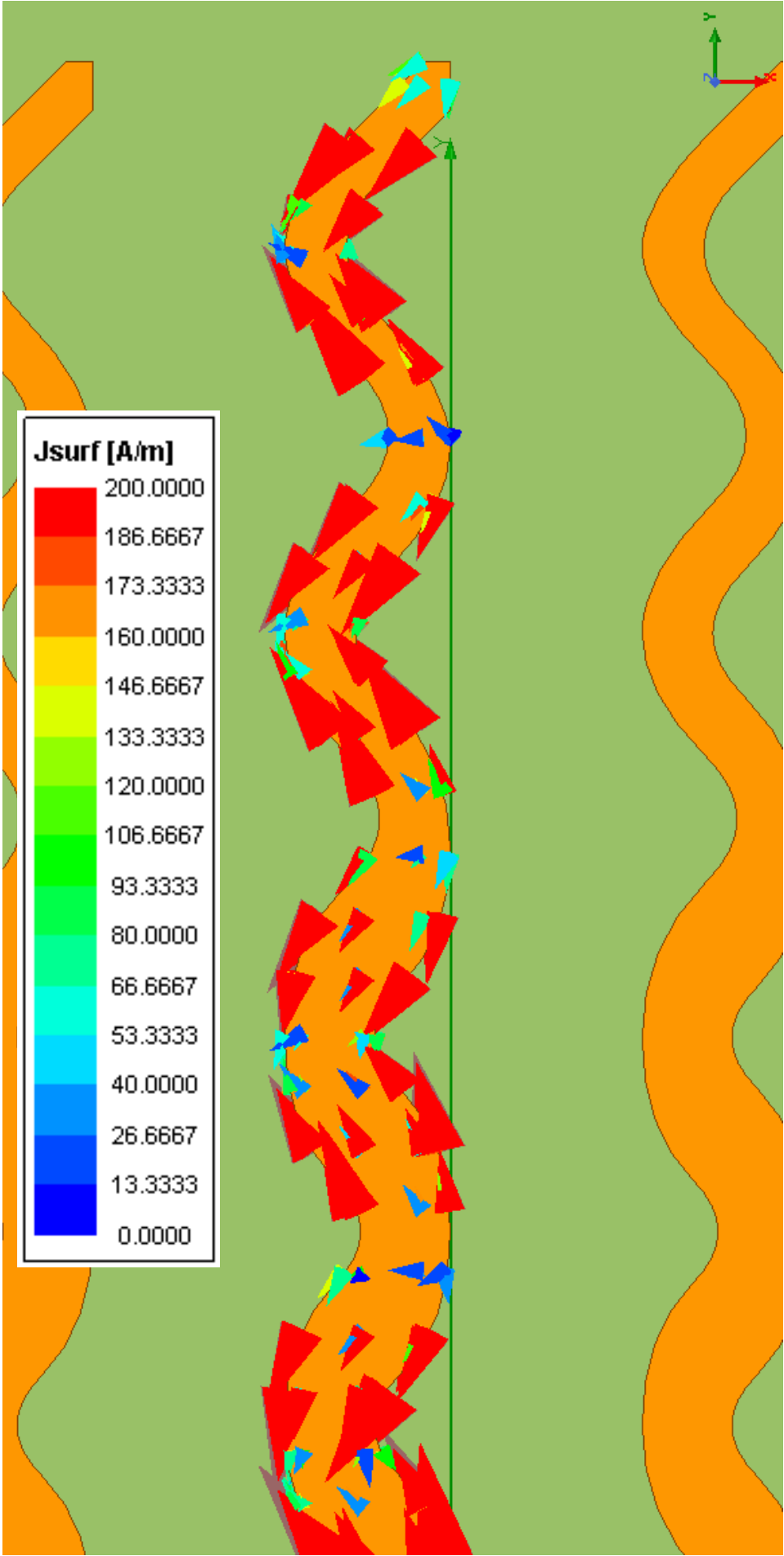}\end{center}

\caption{\emph{Numerical Assessment} ($N=5$, $n=1$) - Screenshot of the
simulated surface current density at $f=f_{0}$. }
\end{figure}

It is worthwhile to point out that the time saving $\Delta t_{sav}$
{[}$\Delta t_{sav}\triangleq\frac{\Delta t_{PSO}-\Delta t_{SbD}}{\Delta t_{PSO}}=\frac{\left(V\times I\right)-\left(S_{0}+S_{upd}\right)}{\left(V\times I\right)}${]}
enabled here by the \emph{SbD} strategy \cite{Massa 2021} with respect
to a standard optimization that exclusively relies on iterated \emph{FW}
calls (i.e., $V\times I$) \cite{Goudos 2016} amounts to $\Delta t_{sav}=85\%$.
As a matter of fact, the \emph{PSO-OK/C} method only requires to simulate
the initial $S_{0}$ training designs and the $S_{upd}$ configurations
adaptively selected during the optimization, while relying on almost
real-time predictions of (\ref{eq:cost-total}) for all the remaining
trial particles generated throughout the iterative minimization procedure
\cite{Massa 2021}. Quantitatively, the synthesis has been completed
in $\Delta t_{SbD}\approx2.25\times10^{2}$ {[}hours{]}, while $\Delta t_{PSO}=1.5\times10^{3}$
{[}hours{]} would be the overall time expense for a standard (non-\emph{SbD})
\emph{PSO}-based optimization %
\footnote{\noindent The average time cost of a single \emph{FW} simulation of
the finite-size model in Fig. 5 is equal to $\Delta t_{FW}\approx45$
{[}min{]} on a desktop \emph{PC} with Intel(R) Core(TM) i7-4790 \emph{CPU}
@ 3.60 {[}GHz{]} and 32 {[}GB{]} of \emph{RAM} memory.%
}.

\section{Performance Assessment\label{sec:Performance-Assessment}}

\noindent The goal of this section is twofold. On the one hand, to
present the results of a careful assessment of the \emph{FW}-simulated
\emph{EM} features of the synthesized \emph{SS-EFA} (Sect. IV-A).
On the other hand, to show the outcomes from the experimental validation,
carried out on a \emph{PCB}-manufactured prototype, to confirm the
\emph{FW} simulations (Sect. IV-B) towards the use of the proposed
\emph{mm}-wave radiator in automotive radar applications.%
\begin{figure}
\begin{center}\includegraphics[%
  width=0.75\columnwidth]{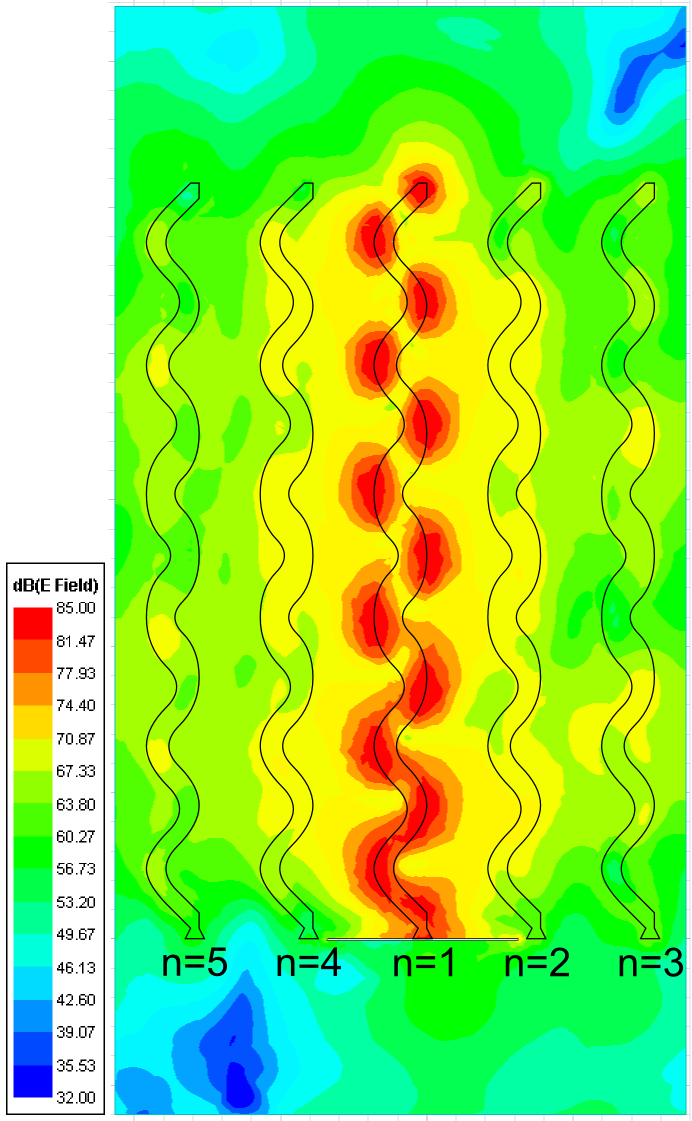}\end{center}

\caption{\emph{Numerical Assessment} ($N=5$, $n=1$) - Screenshot of the
magnitude of the total electric field, $\left|\mathbf{E}\left(x,\, y;\, f_{0}\right)\right|$,
computed over a plane parallel to the $\left(x,\, y\right)$ plane
far $z=\frac{\lambda_{0}}{20}$ from the \emph{SS-EFA} surface, at
$f=f_{0}$.}
\end{figure}
\begin{figure}
\begin{center}\includegraphics[%
  width=0.90\columnwidth]{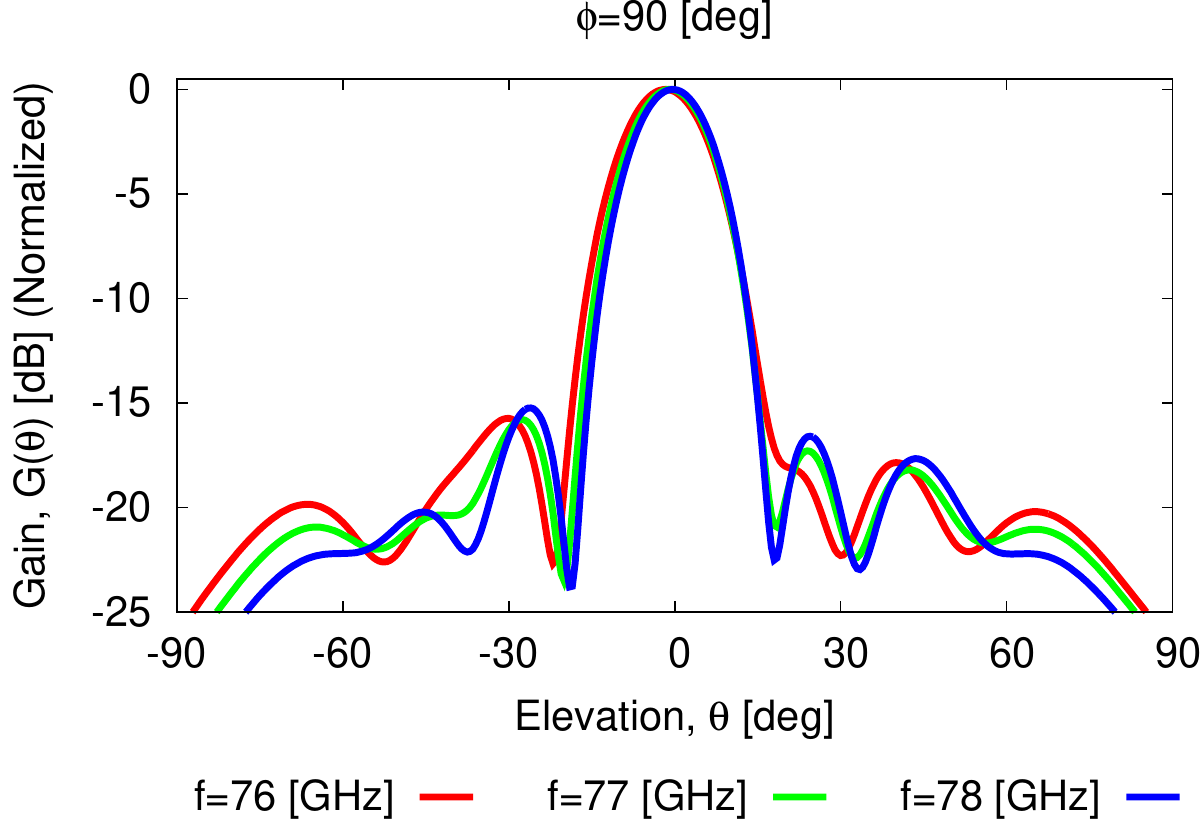}\end{center}

\caption{\emph{Numerical Assessment} ($N=5$, $n=1$, $\varphi=90$ {[}deg{]})
- Embedded gain pattern at $f=f_{\min}$, $f=f_{0}$, and $f=f_{\max}$.}
\end{figure}

\subsection{Numerical Assessment \label{sub:Numerical-Assessment}}

\noindent Figure 6 shows the behavior of the reflection coefficient
of the central ($n=1$) embedded element within the linear arrangement
of $N=5$ identical \emph{SS-EFA}s simulated by means of the Ansys
\emph{HFSS} \emph{FW} solver \cite{HFSS 2021} (Fig. 5). As it can
be inferred, the antenna correctly resonates within the operative
band being $\left|S_{11}\left(f\right)\right|\leq-11.6$ {[}dB{]}
for $f\in\Delta f$. Moreover, the optimized \emph{SS-EFA} provides
a suitable inter-element isolation as indicated by the magnitude of
the scattering coefficients {[}i.e., $\left|S_{n1}\left(f\right)\right|\leq-22.6$
{[}dB{]} ($n=2,...,N$){]}. 

\begin{figure}
\begin{center}\includegraphics[%
  width=0.90\columnwidth]{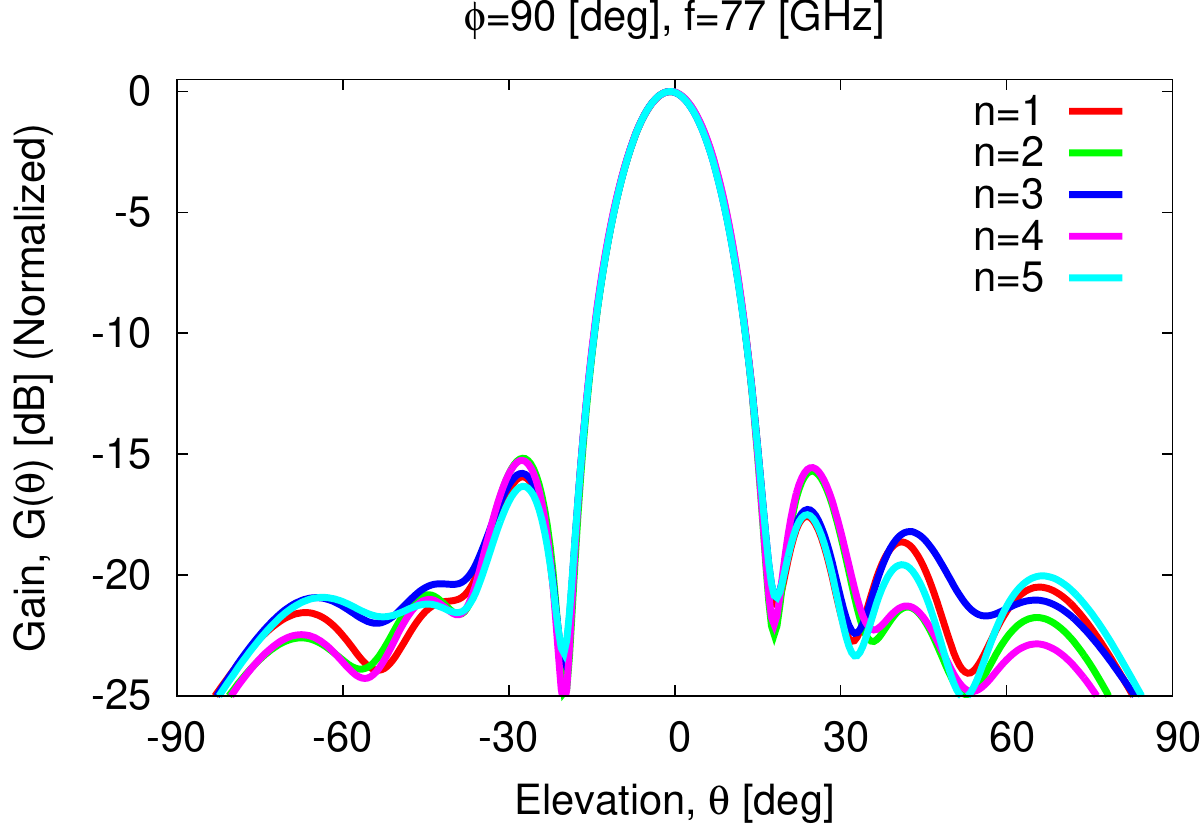}\end{center}

\caption{\emph{Numerical Assessment} ($N=5$, $\varphi=90$ {[}deg{]}, $f=f_{0}$)
- Comparison of the embedded gain patterns of all elements ($n=1,...,N$)
forming the linear arrangement in Fig. 5.}
\end{figure}
As for the \emph{FF} features of the \emph{SbD}-layout, the curves
in Fig. 7(\emph{a}) show that $SLL\left(f\right)\leq-15.3$ {[}dB{]}
and $HPBW\left(f\right)\leq17.5$ {[}deg{]} for $f\in\Delta f$ (i.e.,
both \emph{SLL} and \emph{HPBW} comply with the requirement). Moreover,
the synthesized radiator turns out to be fully-compliant in terms
of beam pointing stability regardless of the adopted edge feeding
mechanism as confirmed by the \emph{BDD} values (i.e., $BDD\leq1.5$
{[}deg{]}) in Fig. 7(\emph{b}). 

On the other hand, it should be very interesting for the readers to
observe that the arising \emph{SS-EFA} structure radiates a linearly
(horizontal) polarized field with high polarization purity despite
its smooth/non-uniform profile (Fig. 1). Indeed, it turns out that
$PR\left(f\right)\geq21$ {[}dB{]} for $f\in\Delta f$ {[}Fig. 7(\emph{b}){]}
as a consequence of the distribution of the surface current (Fig.
8) that follows the theoretically-expected configuration sketched
in Fig. 4. %
\begin{figure}
\begin{center}\includegraphics[%
  width=0.90\columnwidth]{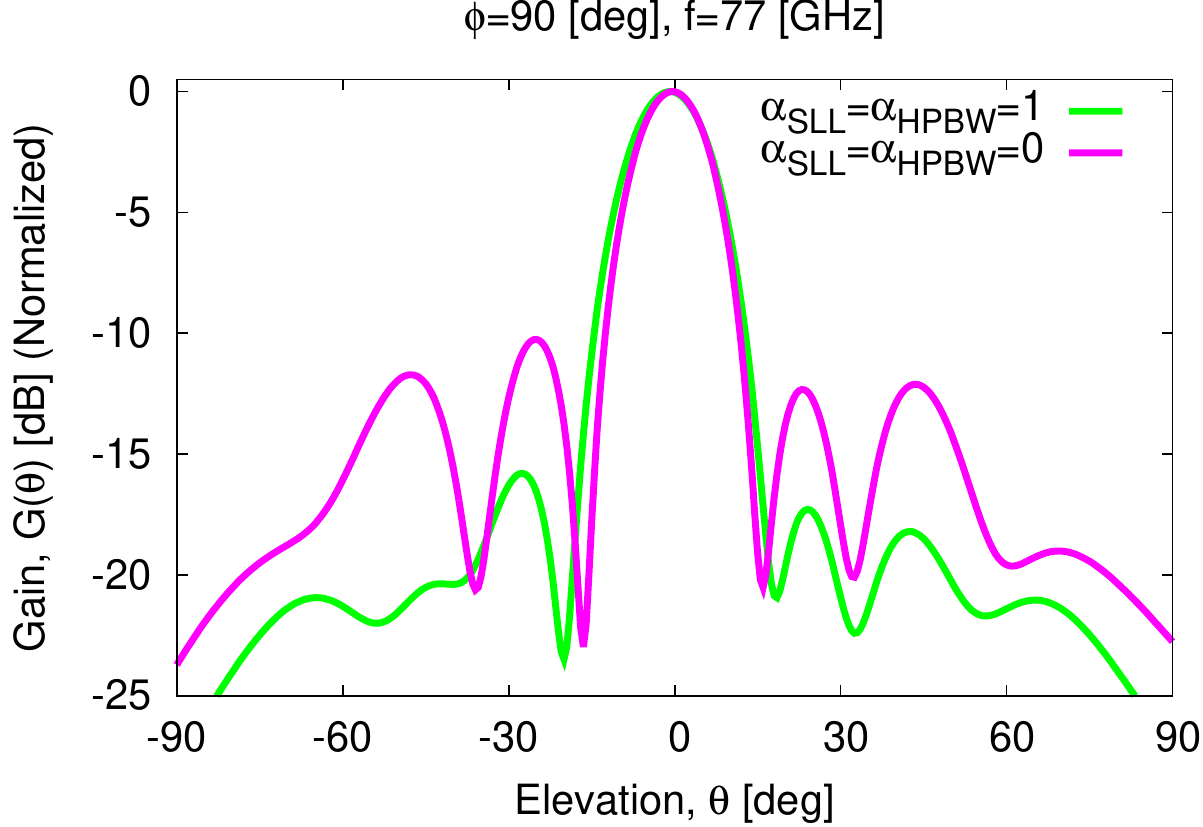}\end{center}

\caption{\emph{Numerical Assessment} ($N=5$, $n=1$, $\varphi=90$ {[}deg{]},
$f=f_{0}$) - Comparison of the embedded gain pattern when enabling
($\alpha_{SLL}=\alpha_{HPBW}=1.0$) or disabling ($\alpha_{SLL}=\alpha_{HPBW}=0.0$)
the optimization of the \emph{SLL} and the \emph{HPBW}.}
\end{figure}
An overall $x$-polarized source is excited in correspondence with
each radiating location of the resonant structure shown in Fig. 9
where the near-field (\emph{NF}) distribution of the magnitude of
the electric field, $\left|\mathbf{E}\left(x,\, y;\, f_{0}\right)\right|$,
computed on a $\left(5W\times L\right)$-sized plane at height $z=\frac{\lambda_{0}}{20}$
from the \emph{SS-EFA}, is reported. This latter plot highlights the
resonant behavior of the synthesized structure, the field being maximum
at fixed and equally-spaced positions. As expected, such maxima, which
correspond to those of the \emph{SW} excited within the structure,
arise on the bends of the spline contour and they generate in-phase
radiation contributions that result in a suitable beam shaping and
a pointing stability within the working band $\Delta f$ as confirmed
by the simulated \emph{FF} pattern at $f=f_{\min}$, $f=f_{0}$, and
$f=f_{\max}$ in Fig. 10. Moreover, the magnitude of the coupling
field in the adjacent radiators ($n=2$ and $n=4$) is always $14.8$
{[}dB{]} lower than in the driven element ($n=1$), the arising \emph{MC}
being always acceptable as indicated by the corresponding scattering
coefficients (i.e., $\left|S_{21}\left(f_{0}\right)\right|=-23.1$
{[}dB{]} and $\left|S_{41}\left(f_{0}\right)\right|=-23.2$ {[}dB{]}
- Fig. 6). For completeness, the embedded pattern of the central element
is compared to that of the (non-optimized) neighboring elements ($n=2,...,N$)
in Fig. 11. As expected, some differences between the elements can
be observed because of the small dimension of the simulated array
(Fig. 5). However, it turns out that such discrepancies are quite
limited thanks to the low coupling among adjacent radiators, thus
further indicating the suitability of the designed radiator for a
successive integration within \emph{FMCW} radars.%
\begin{figure}
\begin{center}\begin{tabular}{c}
\includegraphics[%
  width=0.90\columnwidth]{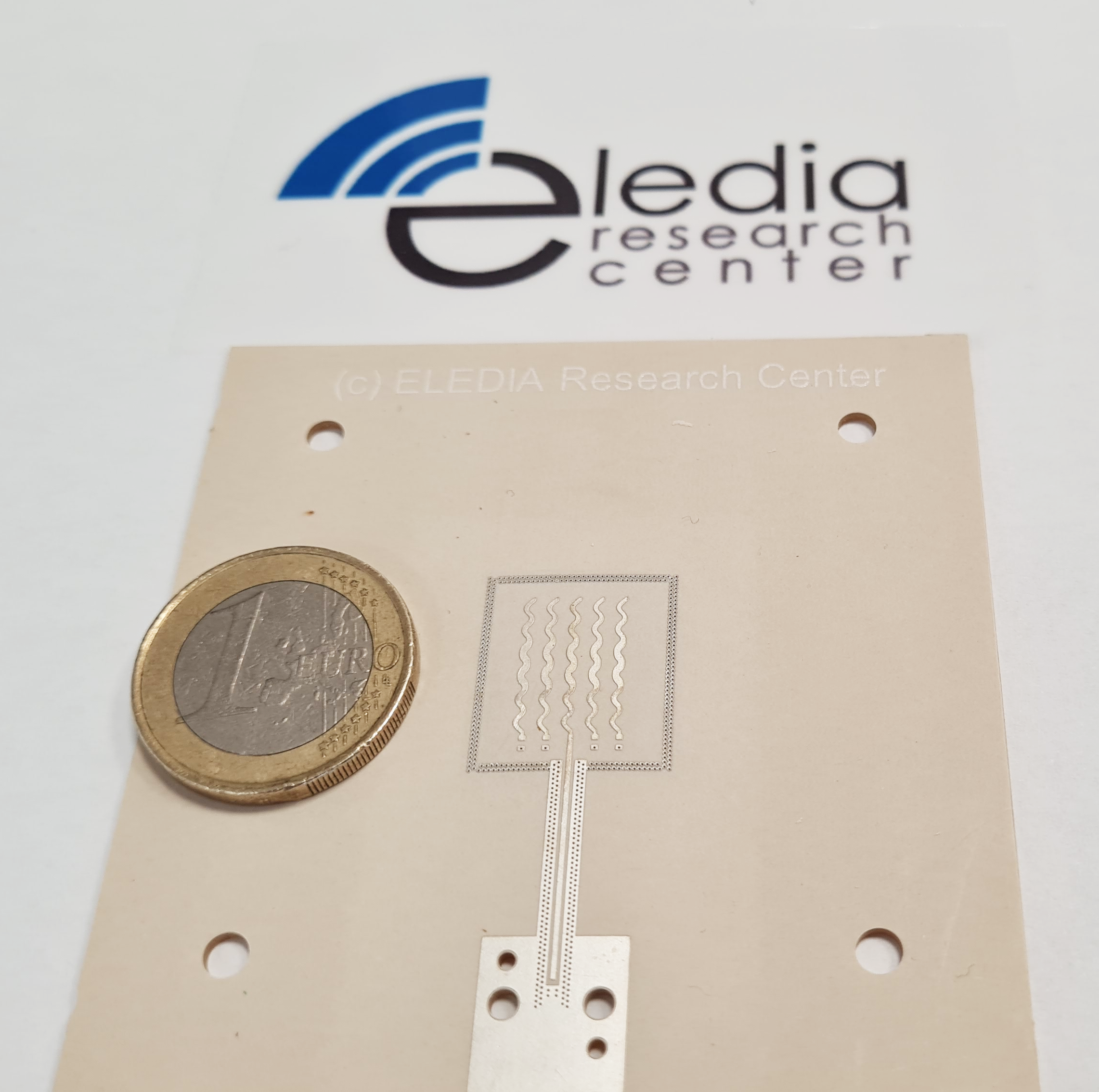}\tabularnewline
(\emph{a})\tabularnewline
\tabularnewline
\includegraphics[%
  width=0.90\columnwidth]{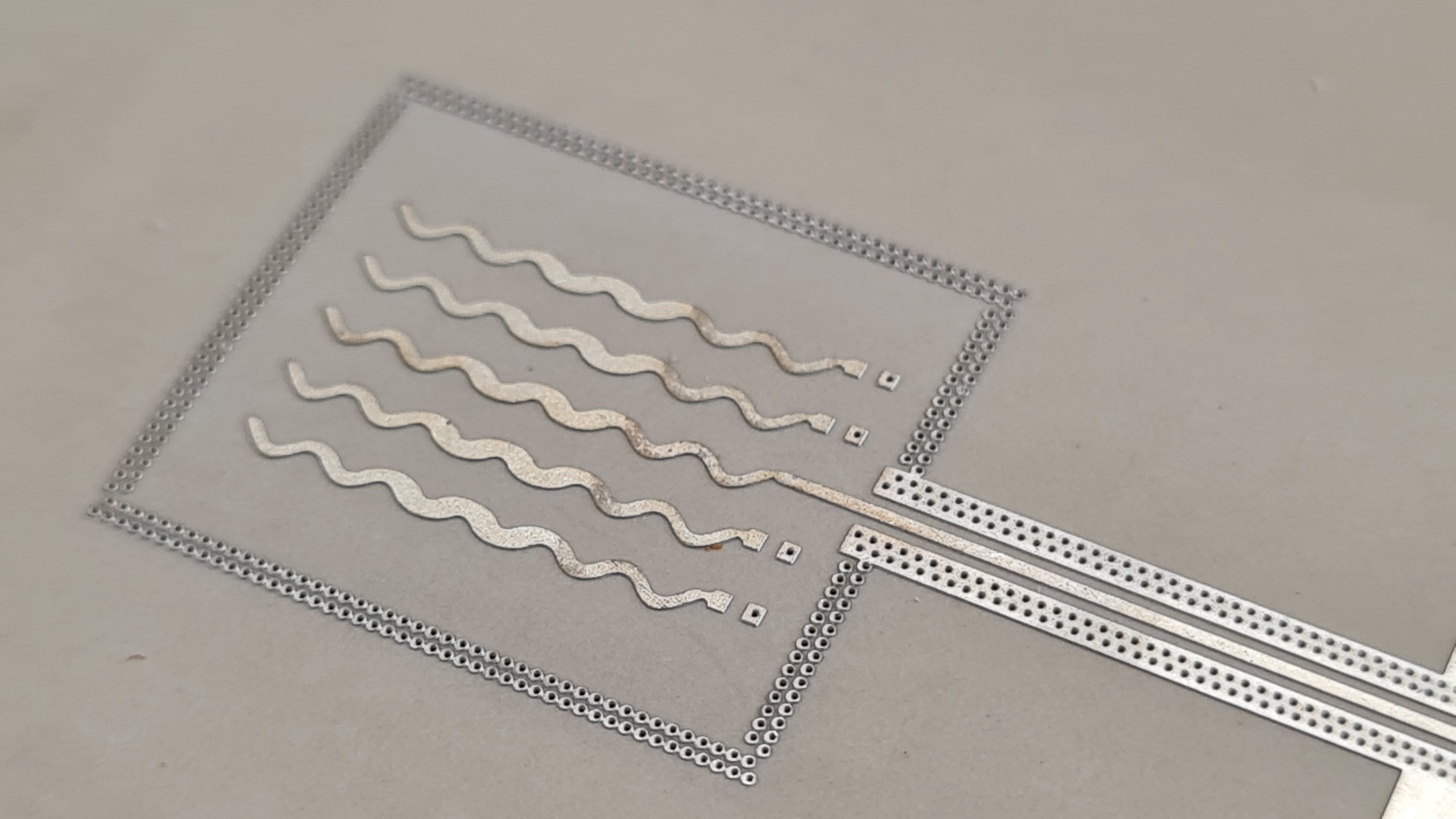}\tabularnewline
(\emph{b})\tabularnewline
\end{tabular}\end{center}

\caption{\emph{Experimental Assessment} - Picture of (\emph{a}) the \emph{SS-EFA}
prototype and (\emph{b}) a zoom on the radiating part.}
\end{figure}

In order to further demonstrate the beam-shaping capabilities of the
proposed optimization strategy, let us consider the case of an antenna
design optimized without enforcing constraints on the \emph{SLL} and
\emph{HPBW}. Towards this end, a second optimization has been performed
by letting the two weights associated to such indexes equal to $\alpha_{SLL}=\alpha_{HPBW}=0.0$,
thus {}``disabling'' the corresponding cost terms (\ref{eq:cost_SLL})(\ref{eq:cost_HPBW})
in the cost function (\ref{eq:cost-total}). The result is shown in
Fig. 12 in terms of pattern of the central element at $f=f_{0}$.
As it can be observed, the absence of specific constraints resulted
in remarkably higher side-lobes (i.e., $\left.SLL\left(f_{0}\right)\right|_{\alpha_{SLL}=\alpha_{HPBW}=0.0}=-10.2$
{[}dB{]} vs. $\left.SLL\left(f_{0}\right)\right|_{\alpha_{SLL}=\alpha_{HPBW}=1.0}=-15.8$
{[}dB{]} - Fig. 12). As a consequence, a narrower beamwidth has been
yielded (i.e., $\left.HPBW\left(f_{0}\right)\right|_{\alpha_{SLL}=\alpha_{HPBW}=0.0}=15.0$
{[}deg{]} vs. $\left.HPBW\left(f_{0}\right)\right|_{\alpha_{SLL}=\alpha_{HPBW}=1.0}=16.5$
{[}deg{]} - Fig. 12). On the other hand, the same beam pointing accuracy
has been kept (i.e., $\left.BDD\right|_{\alpha_{SLL}=\alpha_{HPBW}=0.0}\leq1.5$
{[}deg{]}).%
\begin{table*}

\caption{Comparison between the proposed antenna and other designs with single-layer
structure and horizontal polarization in the recent literature. Gain,
\emph{SLL}, \emph{HPBW}, \emph{PR}, and radiation efficiency are reported
at $f=f_{0}$.}

\begin{center}\resizebox{\textwidth}{!}{\begin{tabular}{|c|c|c|c|c|c|c|c|c|c|c|}
\hline 
Ref.&
Radiator&
Feeding&
$f_{0}$&
\emph{FBW}&
\emph{BDD}&
Gain&
\emph{SLL}&
\emph{HPBW}&
\emph{PR}&
Length\tabularnewline
&
Type&
&
{[}GHz{]}&
{[}\%{]}&
{[}deg{]}&
{[}dBi{]}&
{[}dB{]}&
{[}deg{]}&
{[}dB{]}&
{[}$\lambda_{0}${]}\tabularnewline
\hline
\hline 
This Work&
Spline-Shaped&
\emph{EFA}&
$77.0$&
$2.6$&
$\leq1.5$&
$13.8$&
$-15.8$&
$16.5$&
$22.7$&
$4.9$\tabularnewline
\hline
\hline 
\cite{Joseph 2023}&
Microstrip Line&
\emph{EFA}&
$28.0$&
$4.6$&
N.A.&
$10.7$&
$-15.8$&
$16$&
$18.4$&
$6.0$\tabularnewline
\hline 
\cite{Yi 2023}&
Asymmetric Trapezoidal Microstrip&
\emph{EFA}&
$78.0$&
$13.0$&
$\leq2.2$&
$10.4$&
$-16.1$&
$22.7$&
N.A.&
$3.4$\tabularnewline
\hline 
\cite{Cheng 2009}&
Slotted \emph{SIW}&
\emph{EFA}&
$79.0$&
$5.4$&
\emph{N.A.}&
$6.5$&
$-24.0$&
$34.0$&
$30.1$&
$2.7$\tabularnewline
\hline 
\cite{Ahmed 2020}&
Periodic Microstrip \emph{LWA}&
\emph{EFA}&
$22.5$&
$69.6$&
$\leq55.0$&
$12.9$&
$-13.3$&
$10.5$&
N.A.&
$5.3$\tabularnewline
\hline
\cite{Zhang 2011}&
Linear Grid Array&
\emph{CFA}&
$24.2$&
$1.2$&
\emph{N.A.}&
$19.0$&
$-20.0$&
$5.0$&
$40.0$&
$11.6$\tabularnewline
\hline
\cite{Xu 2014}&
Slotted \emph{SIW}&
\emph{CFA}&
$24.0$&
$1.7$&
$0.0$&
$24.0$&
$-24.5$&
$4.6$&
\emph{N.A.}&
$15.6$\tabularnewline
\hline
\cite{Guo 2020}&
Microstrip Array&
\emph{CFA}&
$27.0$&
$50.6$&
N.A.&
$12.5$&
$-28.4$&
$25.0$&
$22$&
$2.7$\tabularnewline
\hline
\end{tabular}}\end{center}
\end{table*}

\begin{figure}
\begin{center}\begin{tabular}{cc}
\includegraphics[%
  width=0.35\columnwidth]{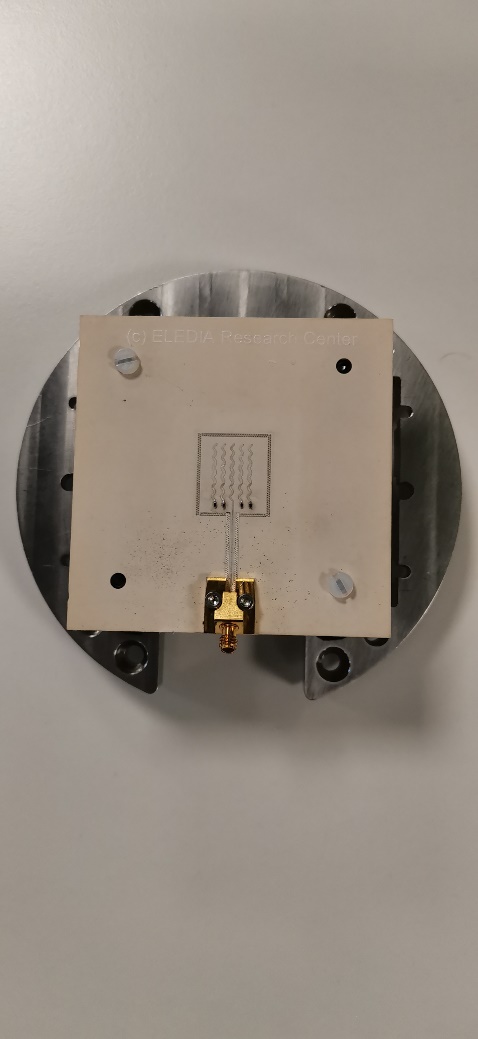}&
\includegraphics[%
  width=0.35\columnwidth]{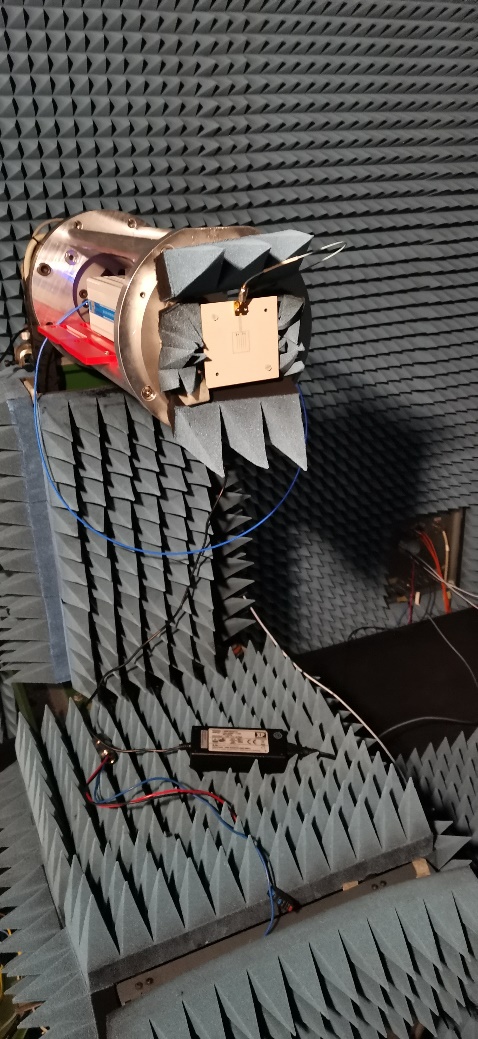}\tabularnewline
(\emph{a})&
(\emph{b})\tabularnewline
&
\tabularnewline
\multicolumn{2}{c}{\includegraphics[%
  width=0.70\columnwidth]{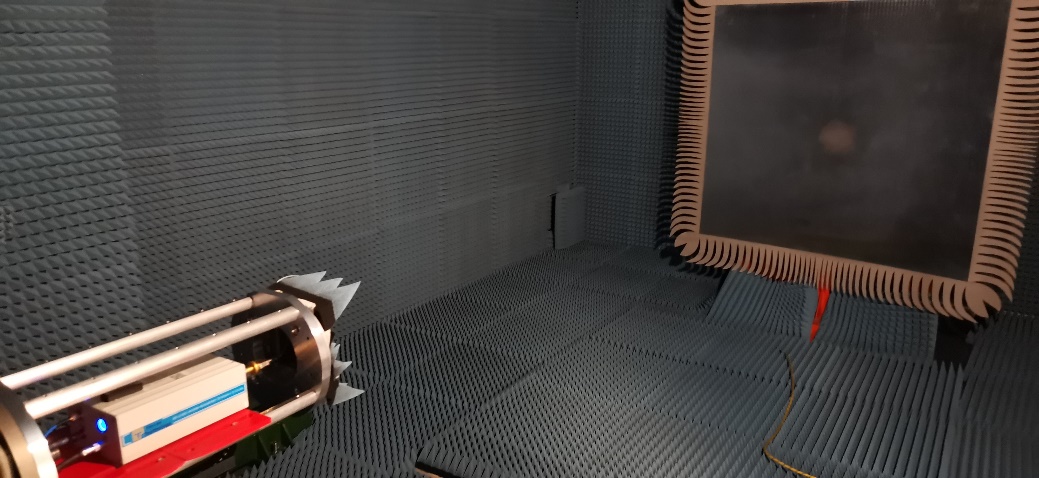}}\tabularnewline
\multicolumn{2}{c}{(\emph{c})}\tabularnewline
\end{tabular}\end{center}

\caption{\emph{Experimental Assessment} - Picture of (\emph{a}) the \emph{AUT},
(\emph{b}) the \emph{CATR} positioner, and (\emph{c}) the whole measurement
setup.}
\end{figure}

Finally, it is worth comparing the proposed antenna with other designs
available in the scientific literature. Towards this end, Table III
summarizes the main features of several state-of-art competitive solutions
complying with the two main requirements of (\emph{a}) horizontal
polarization and (\emph{b}) single-layer layout. As expected, it is
not possible to find a unique {}``winning'' design in terms of all
key performance indicators (\emph{KPI}s), since each technological
solution could in principle represent the best trade-off for a given
application (i.e., project requirements/constraints). However, it
turns out that there are several positive and supporting aspects motivating
the proposal of the antenna at hand. By neglecting the \emph{HPBW}
since (\emph{i}) it is inversely proportional to the length of the
radiating structure (Tab. III) and (\emph{ii}) it is an application-dependent
feature, it appears that the proposed design overcomes state-of-the-art
\emph{EFA} alternatives in several fundamental \emph{KPI}s. For instance,
the performance comparison with the solution in \cite{Joseph 2023}
shows that the proposed design exhibits higher gain ($+29\%$) and
\emph{PR} ($+23\%$) with the same $SLL$. It also outperforms the
design in \cite{Yi 2023} in terms of gain ($+33\%$) and \emph{BDD}
($-32\%$). Moreover, when compared to the solution in \cite{Cheng 2009},
it shows a remarkably higher gain ($+112\%$) with a simpler manufacturing
process not involving vias (as in slotted \emph{SIWs} \cite{Cheng 2009}).
Otherwise, higher gain ($+7\%$) and lower \emph{SLL} ($-19\%$) are
observed with respect to the leaky-wave antenna (\emph{LWA}) in \cite{Ahmed 2020},
which is meant for performing beam scanning with frequency and therefore
providing an unsuitable \emph{BDD} performance for the targeted automotive
application of this work. For the sake of completeness, some \emph{CFA}
solutions have been reported as well (Tab. III). With respect to \cite{Zhang 2011}
and \cite{Xu 2014}, the proposed \emph{SS-EFA} provides a larger
bandwidth ($+117\%$ vs. \cite{Zhang 2011} and $+53\%$ vs. \cite{Xu 2014}),
while it yields a higher gain ($+10\%$) when compared to \cite{Guo 2020}.
Moreover, it is worth remarking that edge-feeding is a highly desirable
feature in radar applications where the radiators must be arranged
into closely-packed/single-layer layouts in which simple routing connections
to the controlling chipset are mandatory. Therefore, the proposed
antenna may be a better (if not the only {}``physically admissible'')
technological choice with respect to such center-fed designs.

\subsection{Experimental Validation \label{sub:Experimental-Assessment}}

\noindent %
\begin{figure}
\begin{center}\includegraphics[%
  width=0.90\columnwidth]{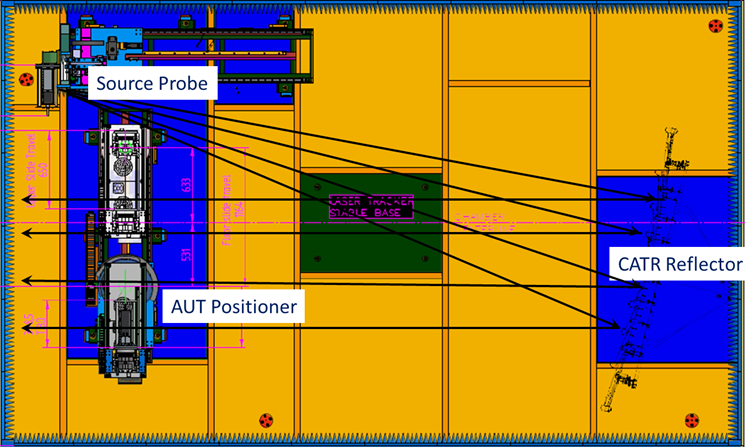}\end{center}

\caption{\emph{Experimental Assessment} - Sketch of the measurement scenario.}
\end{figure}
\begin{figure}
\begin{center}\begin{tabular}{c}
\includegraphics[%
  width=0.70\columnwidth]{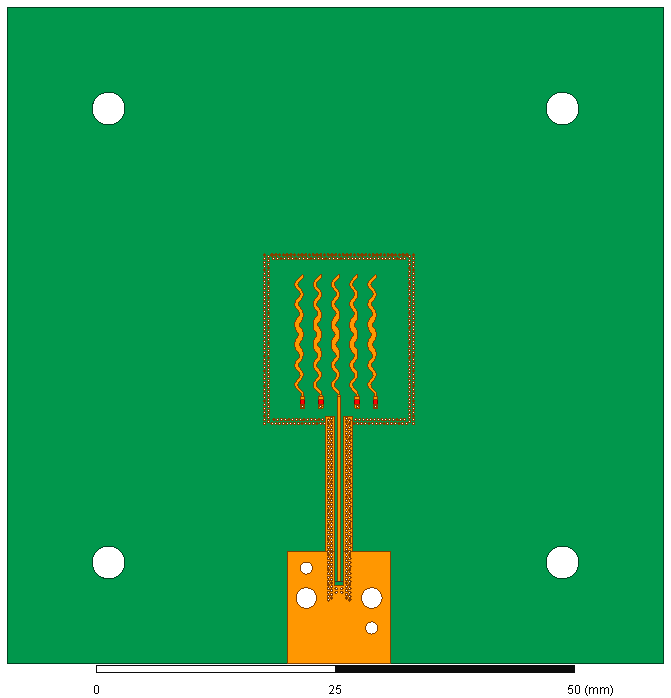}\tabularnewline
(\emph{a})\tabularnewline
\includegraphics[%
  width=0.85\columnwidth]{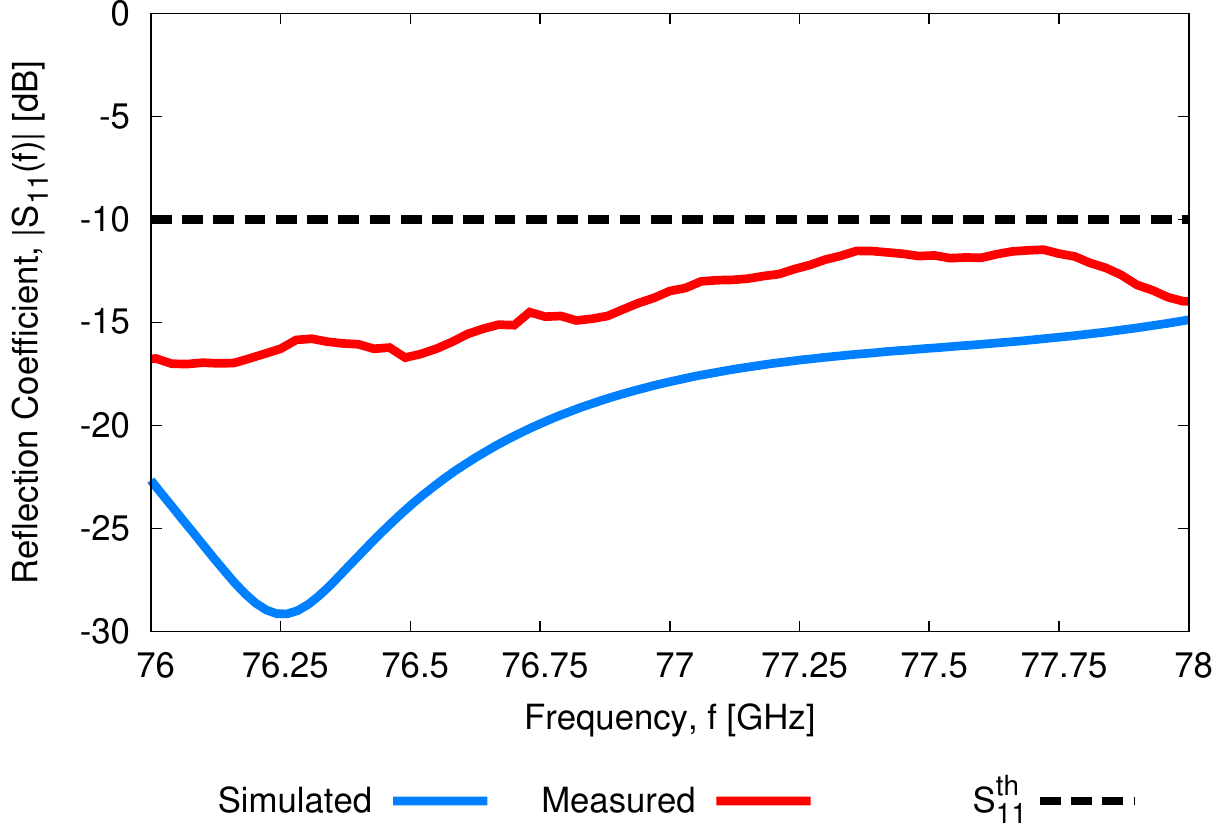}\tabularnewline
(\emph{b})\tabularnewline
\end{tabular}\end{center}

\caption{\emph{Experimental Assessment} - (\emph{a}) Simulated pre-prototyping
layout and (\emph{b}) comparison between simulated and measured input
reflection coefficient of the central element ($n=1$).}
\end{figure}

\noindent A prototype of the \emph{SS-EFA} has been fabricated through
\emph{PCB} manufacturing on a substrate of size $70\times70$ {[}mm{]}
(Fig. 13). The bottom ground plane has been stacked on a $1$ {[}mm{]}-thick
\emph{FR-4} layer to enhance the mechanical robustness and rigidity
of the antenna under test (\emph{AUT}). Moreover, four holes, $3.5$
{[}mm{]} in diameter, have been drilled at the corners of a square
of side $48.5$ {[}mm{]} to fix the \emph{AUT} on the measurement
support by means of plastic screws {[}Fig. 13(\emph{a}) and Fig. 14(\emph{a}){]}.
The central radiator ($n=1$) has been fed by using a Rosemberger
\emph{01K80A-40ML5} connector installed without soldering at the bottom
edge of the \emph{AUT} substrate {[}Fig. 14(\emph{a}){]}. The neighboring
\emph{SS} elements (i.e., $n=2,...,N$) have been terminated on \emph{MCR1}
compact thick film chip resistors (series 0402) acting as matched
loads. Moreover, the active and the dummy radiators as well as the
feeding line have been surrounded by a double set of interleaved vias
to suppress the insurgence of undesired surface currents due to the
electrically-large dimension of the \emph{PCB} {[}Fig. 13(\emph{b}){]}.

\noindent %
\begin{figure}
\begin{center}\begin{tabular}{c}
\includegraphics[%
  width=0.75\columnwidth]{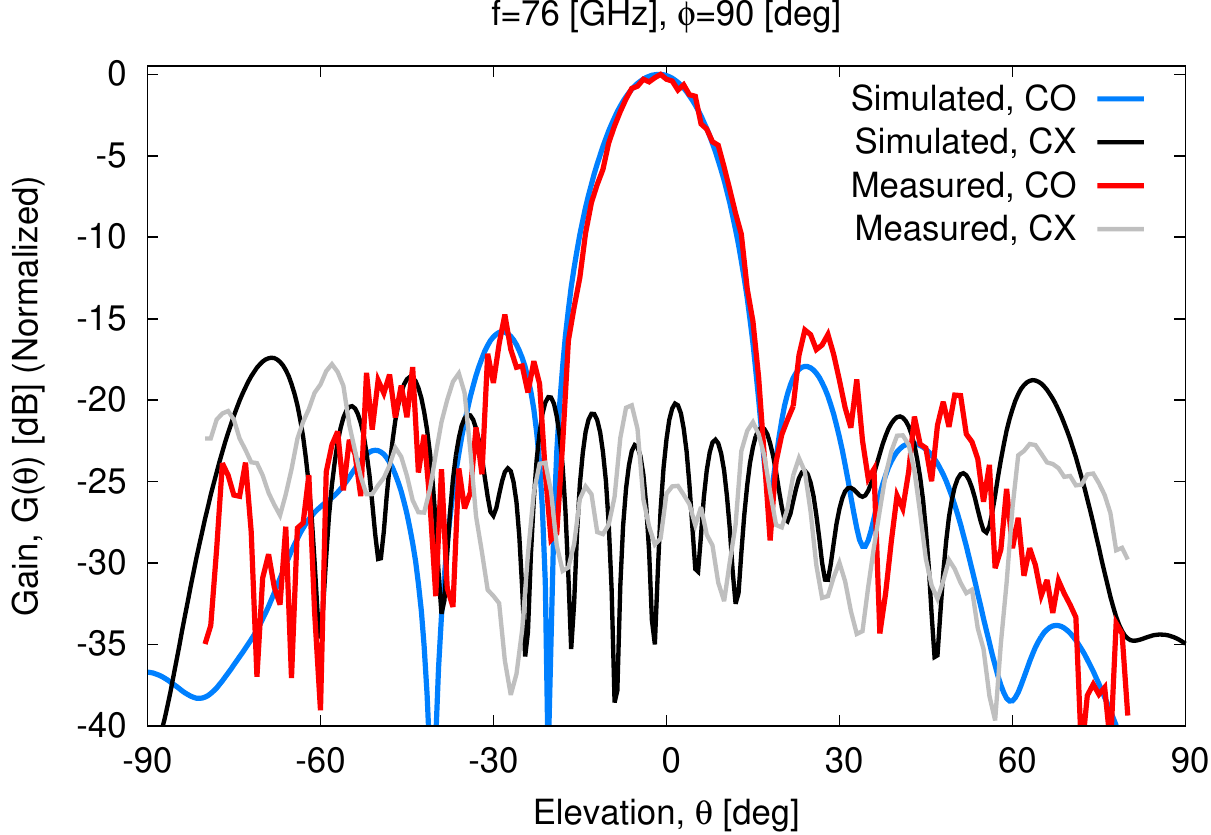}\tabularnewline
(\emph{a})\tabularnewline
\includegraphics[%
  width=0.75\columnwidth]{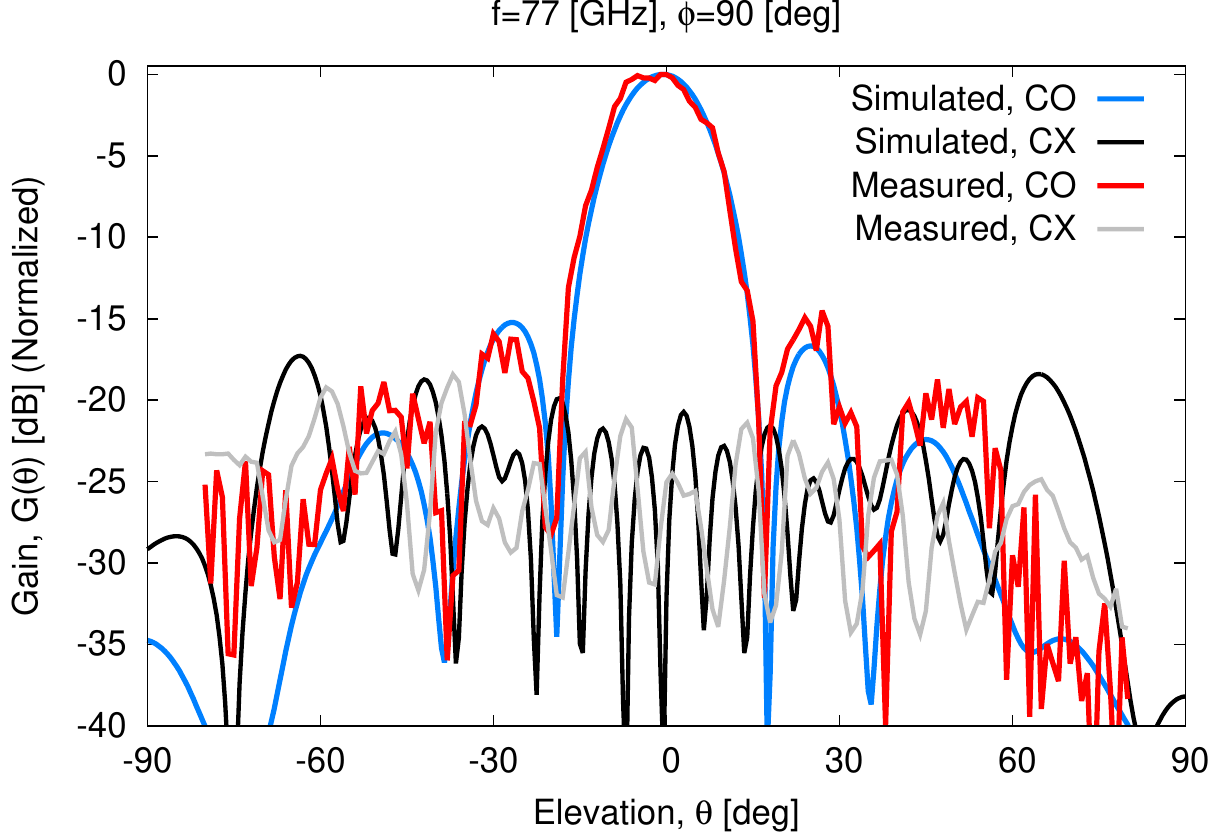}\tabularnewline
(\emph{b})\tabularnewline
\includegraphics[%
  width=0.75\columnwidth]{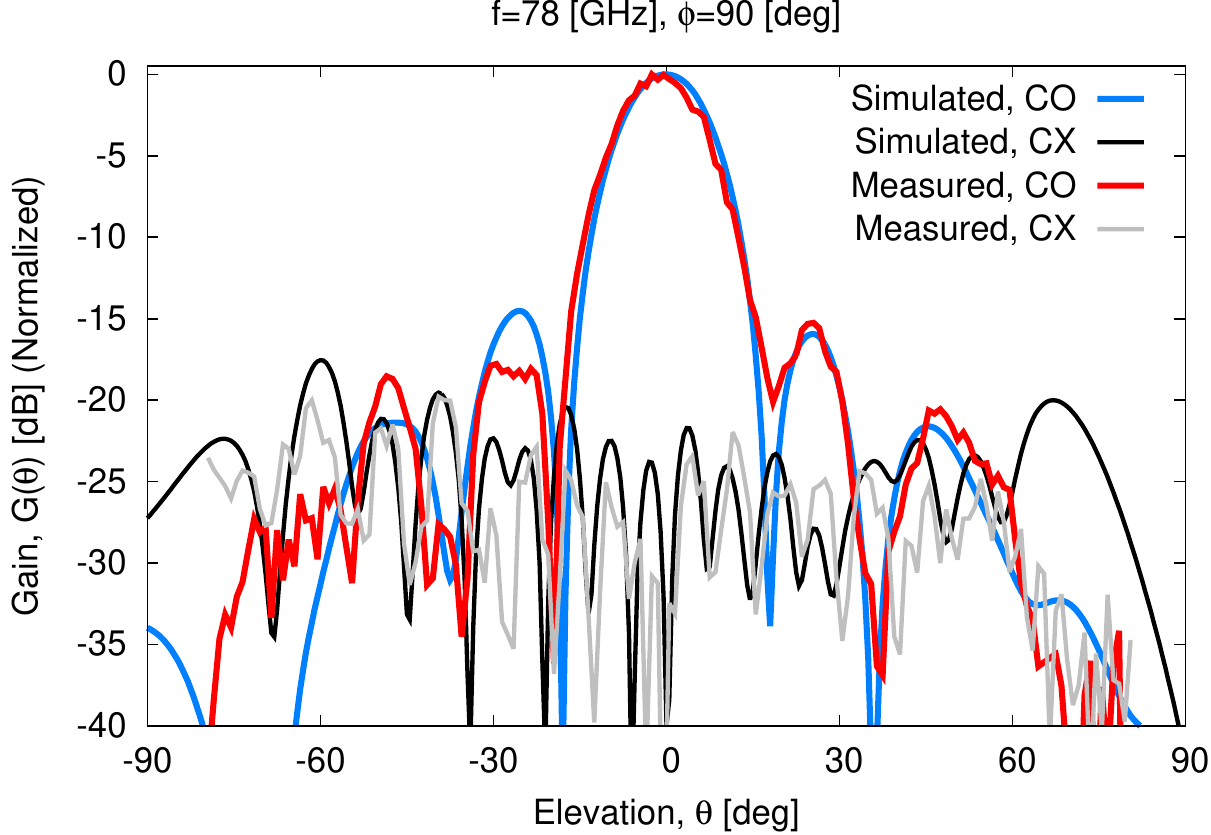}\tabularnewline
(\emph{c})\tabularnewline
\end{tabular}\end{center}

\caption{\emph{Experimental Assessment} ($N=5$, $n=1$, $\varphi=90$ {[}deg{]})
- Comparison between simulated and measured co-polar (\emph{CO}) and
cross-polar (\emph{CX}) embedded element patterns at (\emph{a}) $f=f_{\min}$,
(\emph{b}) $f=f_{0}$, and (\emph{c}) $f=f_{\max}$.}
\end{figure}
The measurement set-up {[}Figs. 14(\emph{b})-14(\emph{c}){]} has been
composed by a \emph{mm}-wave \emph{CATR} system within an Asysol anechoic
chamber (Fig. 15) suitable for antenna measurements up to 170 {[}GHz{]}.

Figure 16(\emph{a}) shows the simulated layout in \emph{Ansys HFSS}
(modeling the prototype of Fig. 13), while a comparison between the
simulated and the measured input reflection coefficient of the central
element ($n=1$) is shown in Fig. 16(\emph{b}). As it can be observed,
the antenna properly resonates in the entire operation band, being
$\left|S_{11}\left(f\right)\right|^{meas}\leq-11.4$ {[}dB{]}, $f\in\Delta f$.
\footnote{\noindent The slight offset of the central frequency in the simulated
$S_{11}$ curve {[}Fig. 16(\emph{b}) vs. Fig. 6{]} is caused by the
modifications to the \emph{PCB} layout in order to include the connector
footprint and the vias {[}Fig. 16(\emph{a}) vs. Fig. 5{]}. %
} Figure 17 shows the \emph{SS-EFA} co-polar (\emph{CO}) and cross-polar
(\emph{CX}) gain patterns measured along the $\varphi=90$ {[}deg{]}-cut
at the minimum {[}$f=f_{\min}$ - Fig. 17(\emph{a}){]}, the central
{[}$f=f_{0}$ - Fig. 17(\emph{b}){]}, and the maximum {[}$f=f_{\max}$
- Fig. 17(\emph{c}){]} operating frequencies. As it can be observed,
the \emph{FF} patterns of the prototype well match the simulated ones\emph{.}
More specifically, there is a very good agreement in both the co-polar
main lobe region and the first left/right sidelobes. Some slight deviations,
occurring especially in the lateral sidelobes, are probably due to
both prototype manufacturing inaccuracies and measurement tolerances
whose impact is certainly significant in the \emph{mm}-wave regime.
Moreover, these experimental results confirm the \emph{FW}-predicted
\emph{SLL} performance since $SLL^{meas}\leq-14.5$ {[}dB{]}, for
$f\in\Delta f$ (Fig. 17). Similarly, the measured \emph{CX} patterns
assess the high polarization purity of the proposed \emph{SS-EFA}
radiating element. As a matter of fact, the normalized \emph{CX} pattern
is always lower than $-18$ {[}dB{]} for $\theta\in\left[-90,\,90\right]$
{[}deg{]} and lower than $-25$ {[}dB{]} along the broadside, for
$f\in\Delta f$ (Fig. 17). Finally, the measured realized gain is
equal to $RG^{meas}\left(f_{\min}\right)=11.8$ {[}dB{]}, $RG^{meas}\left(f_{0}\right)=13.2$
{[}dB{]}, and $RG^{meas}\left(f_{\max}\right)=14.0$ {[}dB{]}, at
the minimum, central, and maximum frequencies, respectively.

\section{Conclusions \label{sec:Conclusion}}

\noindent A novel radiating element for 77 GHz automotive radars has
been proposed that relies on a spline-based modeling to yield a high
geometric flexibility with a reduced number of \emph{DoF}s, while
enabling the fitting of the several contrasting requirements on bandwidth
and \emph{FF} features. The synthesized layout, which has been efficiently
obtained by means of a customized \emph{SbD} approach, provides a
proper input impedance matching, a high isolation, a suitable \emph{SLL}/\emph{HPBW}
as well as a high polarization purity and a stable beam pointing over
frequency regardless of the edge-feeding mechanism. The experimental
assessment, carried out in a \emph{CATR mm}-wave system on a \emph{PCB}-manufactured
prototype, has verified the \emph{FW}-predicted radiation features
over the operative band.

\section*{Acknowledgements}

\noindent A. Massa wishes to thank E. Vico for her never-ending inspiration,
support, guidance, and help.

\begin{IEEEbiography}[{\includegraphics[width=1in,height=1.25in,clip,keepaspectratio]{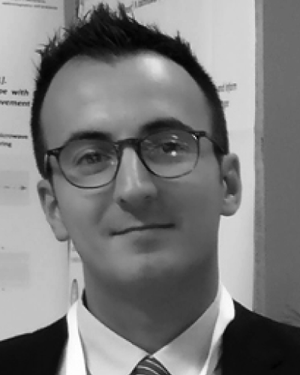}}]{Marco Salucci} (Senior Member, IEEE) received the M.S. degree in Telecommunication Engineering from the University of Trento, Italy, in 2011, and the Ph.D. degree from the International Doctoral School in Information and Communication Technology of Trento in 2014. He was a Post Doc Researcher at CentraleSup\'elec, in Paris, France, and a Post Doc Researcher at the Commissariat à  l'\'Energie Atomique et aux \'Energies Alternatives (CEA), France. He is currently tenure track Associate Professor at the Department of Civil, Environmental, and Mechanical Engineering (DICAM) at the University of Trento, and a Research Fellow of the ELEDIA Research Center. Dr. Salucci is a Member of the IEEE Antennas and Propagation Society and he was a Member of the COST Action TU1208 "Civil Engineering Applications of Ground Penetrating Radar". He is the Associate Editor for Communications and Memberships of the IEEE Transactions on Antennas and Propagation. Moreover, he serves as an Associate Editor of the IEEE Transactions on Antennas and Propagation, the IEEE Open Journal of Antennas and Propagation, and the International Journal of Microwave and Wireless Technologies. Furthermore, he serves as a reviewer for different international journals including the IEEE Transactions on Antennas and Propagation, the IEEE Transactions on Microwave Theory and Techniques, the IEEE Transactions on Geoscience and Remote Sensing, and the IEEE Journal on Multiscale and Multiphysics Computational Techniques. In 2023, he co-edited the book "Applications of Deep Learning in Electromagnetics - Teaching Maxwell's Equations to Machines" published by the IET. His research activities are mainly concerned with inverse scattering, biomedical and GPR microwave imaging techniques, antenna synthesis, and computational electromagnetics with focus on System-by-Design methodologies integrating optimization techniques and artificial intelligence for real-world applications.\end{IEEEbiography}\begin{IEEEbiography}[{\includegraphics[width=1in,height=1.25in,clip,keepaspectratio]{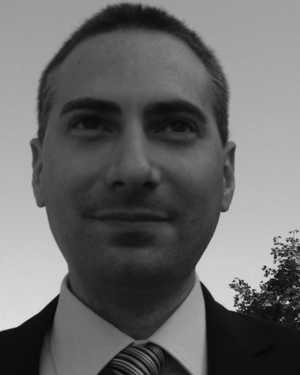}}]{Lorenzo Poli} (Senior Member, IEEE) received the M.S. degree in Telecommunication Engineering from the University of Trento, Italy, in 2008, and the PhD degree from the International Doctoral School in Information and Communication Technology in 2012. He is currently Assistant Professor at the Department of Civil, Environmental, and Mechanical Engineering (DICAM) at the University of Trento, and a Research Fellow of the ELEDIA Research Center. Dr. Poli is author/co-author of more than 60 journals and 90 conference papers. He is a member of the IEEE Antennas and Propagation Society since 2010, when he was a recipient of the IEEE Antennas and Propagation Society Doctoral Research Award. He was a Visiting Researcher at the Laboratoire des Signaux et Systèmes (L2S@Supélec, France) in 2015 and a Visiting Professor at the University of Paris Sud (France) in 2016. Dr. Poli serves as a reviewer for several international journals including IEEE Transactions on Antennas and Propagation, IEEE Antennas and Wireless Propagation Letters, and IET Microwaves, Antennas, and Propagation. His research activities are focused on the solution of antenna design and unconventional array synthesis problems as well as electromagnetic inverse scattering problems.\end{IEEEbiography}\begin{IEEEbiography}[{\includegraphics[width=1in,height=1.25in,clip,keepaspectratio]{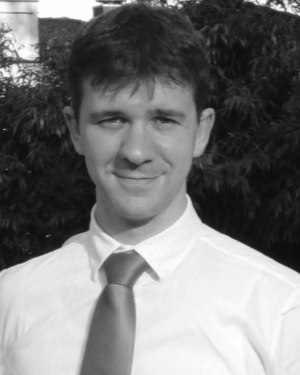}}]{Paolo Rocca} (IEEE Fellow) received the MS degree in Telecommunications Engineering (summa cum laude) in 2005 and the PhD Degree in Information and Communication Technologies in 2008 from the University of Trento, Italy. He is currently Associate Professor at the Department of Civil, Environmental, and Mechanical Engineering (University of Trento), Huashan Scholar Chair Professor at the Xidian University, Xi'an, China, and a member of the ELEDIA Research Center. Moreover, he is Member of the Big Data and AI Working Group for the Committee on Engineering for Innovative Technologies (CEIT) of the World Federation of Engineering Organizations (WFEO). Prof. Rocca received the National Scientific Qualification for the position of Full Professor in Italy and France in April 2017 and January 2020, respectively. Prof. Rocca is author/co-author of 2 book chapters, 165 journal papers, and more than 290 conference papers. He has been a visiting Ph.D. student at the Pennsylvania State University (USA), at the University Mediterranea of Reggio Calabria (Italy), and a visiting researcher at the Laboratoire des Signaux et Systèmes (L2S@ Supélec, France) in 2012 and 2013. Moreover, he has been an Invited Professor at the University of Paris Sud (France) in 2015 and at the University of Rennes 1 (France) in 2017. Prof. Rocca has been awarded from the IEEE Geoscience and Remote Sensing Society and the Italy Section with the best PhD thesis award IEEE-GRS Central Italy Chapter. His main interests are in the framework of artificial intelligence techniques as applied to electromagnetics, antenna array synthesis and analysis, electromagnetic inverse scattering, and quantum computing for electromagnetic engineering. He served as an Associate Editor of the IEEE Antennas and Wireless Propagation Letters (2011-2016), the Microwave and Optical Technology Letters (2019-2020) and serves as an Associate Editor of the IEEE Antennas and Propon Magazine (snce 2020), and  Engineering (since 2020).\end{IEEEbiography}\begin{IEEEbiography}[{\includegraphics[width=1in,height=1.25in,clip,keepaspectratio]{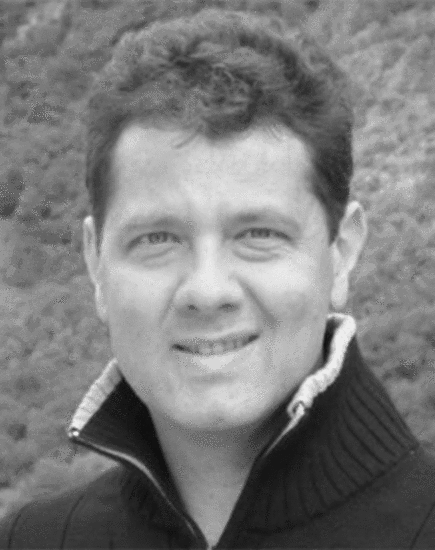}}]{Claudio Massagrande} received the master's degree in electronic engineering (Telecommunications) from Politecnico di Milano, Milan, Italy, in 1996. He is a Principal Antenna Engineer with the Huawei Milan Research Center (MiRC), Milan, Italy. He has almost 25 years of experience in the telecom industry as a microwave and antenna designer for Andrew (now Commscope), Siemens (then Nokia Siemens Networks), and Huawei Technologies. During his carrier, he designed, developed, and tested front ends of several families of microwave radios gaining considerable experience in microwave assembly technologies, such as chip and wire and die attach. His expertise ranges from radio frequency design for microwave PTP radio links for the mobile network backhaul to radio link integration to advanced antenna systems design. His current research activity is focused on advanced antenna systems for mm-wave 5G access.\end{IEEEbiography}\begin{IEEEbiography}[{\includegraphics[width=1in,height=1.25in,clip,keepaspectratio]{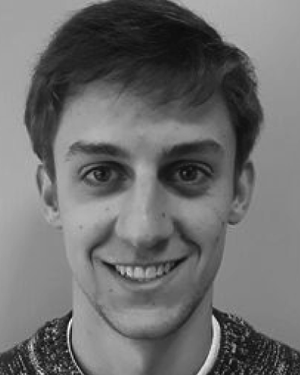}}]{Pietro Rosatti} received the B.Sc. degree in electronics and telecommunications engineering and the M.Sc. degree in information and communication engineering from the University of Trento, Trento, Italy, in 2017 and 2020, respectively, where he is currently pursuing the Ph.D. degree in information and communication technology with the International Doctoral School. He is also a Senior Researcher with the ELEDIA Research Center, University of Trento. His research interests include the design and analysis of wide-angle phased array antennas for modern applications, such as automotive radars and 5G communications.\end{IEEEbiography}\begin{IEEEbiography}[{\includegraphics[width=1in,height=1.25in,clip,keepaspectratio]{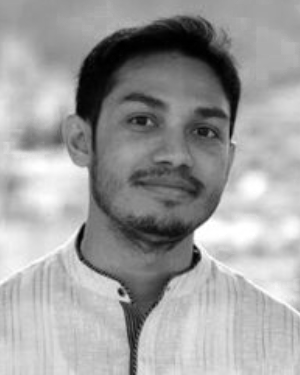}}]{Mohammad Abdul Hannan} received the B. Sc. in Electronics and Telecommunication Engineering (ETE) from Daffodil International University, Bangladesh in 2010, the Master Degree in Telecommunication Engineering from University of Trento, Italy, in 2015, and the Ph.D. degree from the International Doctoral School in Information and Communication Technology, Trento, Italy, in 2020. He is currently  Assistant Professor at the Department of Electrical Electronic and Computer Engineering at the University of Catania, Italy, and a Senior Researcher of the ELEDIA Research Center. His research work is mainly focused on electromagnetic direct and inverse scattering and antenna system synthesis for sensing and communications.\end{IEEEbiography}\begin{IEEEbiography}[{\includegraphics[width=1in,height=1.25in,clip,keepaspectratio]{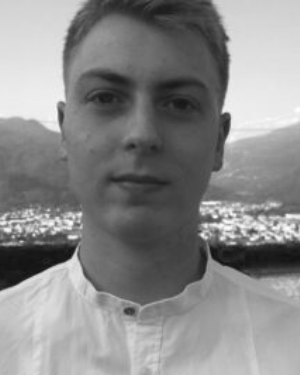}}]{Mirko Facchinelli} received the B.Sc. degree in communication and information engineering and the M.Sc. degree in information and communication engineering from the University of Trento, Trento, Italy, in 2021 and 2023. He is an ETRP-Advanced at the ELEDIA Research Center. His research activity is focused on the optimization techniques for the design of simplified antenna arrays.\end{IEEEbiography}\begin{IEEEbiography}[{\includegraphics[width=1in,height=1.25in,clip,keepaspectratio]{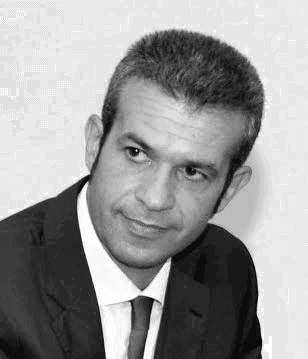}}]{Andrea Massa} (IEEE Fellow, IET Fellow, Electromagnetic Academy Fellow) received the Laurea (M.S.) degree in Electronic Engineering from the University of Genoa, Genoa, Italy, in 1992 and the Ph.D. degree in EECS from the same university in 1996. He is currently a Full Professor of Electromagnetic Fields at the University of Trento, where he currently teaches electromagnetic fields, inverse scattering techniques, antennas and wireless communications, wireless services and devices, and optimization techniques. At present, Prof. Massa is the director of the network of federated laboratories "ELEDIA Research Center" (www.eledia.org) located in Brunei, China, Czech, France, Greece, Italy, Japan, Perù, Tunisia with more than 150 researchers. Moreover, he is holder of a Chang-Jiang Chair Professorship @ UESTC (Chengdu - China), Visiting Research Professor @ University of Illinois at Chicago (Chicago - USA), Visiting Professor @ Tsinghua (Beijing - China), Visiting Professor @ Tel Aviv University (Tel Aviv - Israel), and Professor @ CentraleSupélec (Paris - France). He has been holder of a Senior DIGITEO Chair at L2S-CentraleSupélec and CEA LIST in Saclay (France), UC3M-Santander Chair of Excellence @ Universidad Carlos III de Madrid (Spain), Adjunct Professor at Penn State University (USA), Guest Professor @ UESTC (China), and Visiting Professor at the Missouri University of Science and Technology (USA), the Nagasaki University (Japan), the University of Paris Sud (France), the Kumamoto University (Japan), and the National University of Singapore (Singapore). He has been appointed IEEE AP-S Distinguished Lecturer (2016-2018) and served as Associate Editor of the "IEEE Transaction on Antennas and Propagation" (2011-2014). Prof. Massa serves as Associate Editor of the "International Journal of Microwave and Wireless Technologies" and he is member of the Editorial Board of the "Journal of Electromagnetic Waves and Applons", a permannt member of th"PIERS Technical Committee" and of the "EuMW Technical Committee", and a ESoA member. He has been appointed in the Scientific Board of the "Società  Italiana di Elettromagnetismo (SIEm)" and elected in the Scientific Board of the Interuniversity National Center for Telecommunications (CNIT). He has been appointed in 2011 by the National Agency for the Evaluation of the University System and National Research (ANVUR) as a member of the Recognized Expert Evaluation Group (Area 09, "Industrial and Information Engineering") for the evaluation of the researches at the Italian University and Research Center for the period 2004-2010. Furthermore, he has been elected as the Italian Member of the Management Committee of the COST Action TU1208 "Civil Engineering Applications of Ground Penetrating Radar". His research activities are mainly concerned with inverse problems, analysis\/synthesis of antenna systems and large arrays, radar systems synthesis and signal processing, cross-layer optimization and planning of wireless\/RF systems, semantic wireless technologies, system-by-design and material-by-design (metamaterials and reconfigurable-materials), and theory\/applications of optimization techniques to engineering problems (tele-communications, medicine, and biology). Prof. Massa published more than 900 scientific publications among which more than 350 on international journals (> 15.000 citations  h-index = 65 [Scopus]; > 12.000 citations  h-index = 59 [ISI-WoS]; > 23.000 citations  h-index = 89 [Google Scholar]) and more than 550 in international conferences where he presented more than 200 invited contributions (> 40 invited keynote speaker) (www.eledia.org/publications). He has organized more than 100 scientific sessions in international conferences and has participated to several technological projects in the national and international framework with both national agencies and companies (18 international prj, > 5 MEu; 8 national prj, > 5 MEu; 10 local prj, > 2 MEu; 63 industrial prj, > 10 MEu; 6 university prj, > 300 KEu).\end{IEEEbiography}

\EOD
\end{document}